\documentclass{ptephy_v1}
\usepackage{latexsym,graphicx,amssymb,amsmath,mathrsfs}
\usepackage{multirow}
\usepackage{epstopdf}
\usepackage{url}
\graphicspath{{./Figs/}}
\newcommand{\LL}{\Lambda\Lambda}
\newcommand{\nuc}[2]{{}^{#1}\mathrm{#2}}
\newcommand{\nucL}[2]{{}^{#1}_\Lambda\mathrm{#2}}
\newcommand{\nucLL}[2]{{}^{#1}_{\LL}\mathrm{#2}}
\newcommand{\HeLL}[1]{\nucLL{\,\;\;{#1}}{He}}
\newcommand{\HLL}[1]{\nucLL{\,\;\;{#1}}{H}}
\newcommand{\Lambpha}{\HeLL{6}}
\newcommand{\nucX}[2]{{}^{#1}_{\Xi}\mathrm{#2}}
\newcommand{\dBLL}{\Delta B_{\LL}}
\newcommand{\MeV}{\mathrm{MeV}}
\newcommand{\DLC}{D$\Lambda$C}
\newcommand{\DLH}{D$\Lambda$HN}
\newcommand{\hyph}{\operatorname{-}}
\newcommand{\PDLC}{P_\text{\DLC}}
\newcommand{\PDLH}{P_\text{\DLH}}
\newcommand{\PDLHtot}{P_\text{\DLH}^\mathrm{tot}}
\newcommand{\Pbr}{P_\mathrm{Br}}
\newcommand{\PbrTot}{\Pbr^\mathrm{tot}}
\newcommand{\Nev}{N_\mathrm{ev}}
\usepackage[normalem]{ulem}  

\newcommand{\ignore}[1]{}

\renewcommand\sout{\bgroup \color{red} \ULdepth=-.5ex \ULset}

\begin{document}

\preprintnumber{YITP-19-105}

\title{Statistical double $\Lambda$ hypernuclear formation
from $\Xi^-$ absorption at rest in light nuclei\footnote{
Report number: YITP-19-105}}

\author[1]{Akira Ohnishi}
\affil{Yukawa Institute for Theoretical Physics, Kyoto University,
Kyoto 606-8502, Japan
\email{ohnishi@yukawa.kyoto-u.ac.jp}}

\author[2]{Chikako Ishizuka}
\affil{Laboratory for Advanced Nuclear Energy, Institute of Innovative Research,
Tokyo Institute of Technology, Tokyo 152-8550, Japan}

\author[2,3]{Kohsuke Tsubakihara}
\affil{National Institute of Technology, Asahikawa Colledge,
Asahikawa 071-8142, Japan}

\author[4]{Yuichi Hirata}
\affil{
Central Institute of Isotope Science,
Hokkaido University,
Sapporo 060-0815, Japan}

\begin{abstract}
We investigate double $\Lambda$ hyperfragment formation
from the statistical decay of double $\Lambda$ compound nuclei
produced in the $\Xi^-$ absorption at rest in light nuclei,
$^{12}\mathrm{C}$, $^{14}\mathrm{N}$ and $^{16}\mathrm{O}$.
We examine the target and the $\Lambda\Lambda$ bond energy dependence
of the double $\Lambda$ hyperfragment formation probabilities,
especially of those double hypernuclei observed in experiments.
For the $^{12}\mathrm{C}$ ($^{14}\mathrm{N}$) target,
the formation probabilities of 
$^{\,\;\;6}_{\Lambda\Lambda}\mathrm{He}$
and 
$^{\;10}_{\Lambda\Lambda}\mathrm{Be}$
($^{\;13}_{\Lambda\Lambda}\mathrm{B}$)
are found to be reasonably large
as they are observed in the KEK-E373 (KEK-E176) experiment.
By comparison, for $^{16}\mathrm{O}$ target,
the formation probability of $^{\;11}_{\Lambda\Lambda}\mathrm{Be}$
is calculated to be small 
with $\Delta B_{\Lambda\Lambda}$ consistent with the Nagara event.
We also evaluate the formation probability
of ${}^{\,\;\;5}_{\Lambda\Lambda}\mathrm{H}$
from a $\Xi^-$-${}^{6}\mathrm{He}$ bound state, ${}^{7}_{\Xi}\mathrm{H}$.
\end{abstract}

\maketitle

\section{Introduction}

Formation of double $\Lambda$ hypernuclei (\DLH) from $\Xi^-$ absorption
in nuclei is of importance in several aspects.
The $\Xi^-$ absorption at rest in nuclei is the most efficient way
to produce \DLH,
and uniquely identified \DLH~\cite{Danysz1963,E176,Nagara,E373,E07}
provide strong constraints on the $\LL$
interaction~\cite{Hiyama2010,FG2002}.
The strength and density dependence of the $\LL$ interaction
are the keys to solve the hyperon puzzle in neutron star physics.
Until now, four \DLH\ formation events
have been uniquely identified,
and more will be found in the J-PARC-E07 experiment,
where $10^4$ $\Xi^-$ absorption events in nuclei are expected to be observed.
Let us comment on these points in order.

The baryon-baryon interaction has been one of the central subjects
in nuclear physics.
Compared with the nucleon-nucleon interaction,
hyperon-nucleon scattering data are much more scarce
and single $\Lambda$ hypernuclear data are also used to constrain
the $\Lambda N$ interaction.
For the $\LL$ interaction,
there are theoretical predictions
in the meson exchange model~\cite{ESC08},
the quark cluster model~\cite{fss2},
and the lattice QCD calculations~\cite{Sasaki:2015ifa}.
Experimentally, by comparison,
it is not possible to perform scattering experiments,
and hence the binding energies of
\DLH~\cite{Danysz1963,E176,Nagara,E373,E07}
and the correlation function data from high-energy nuclear
collisions~\cite{CorrExp,CorrTheory}
have been utilized to experimentally constrain the $\LL$ interaction.
While the correlation function technique is recently applied
to get knowledge on several hadron-hadron interactions~\cite{CorrExp},
we need further theoretical and experimental studies
to constrain interactions precisely~\cite{CorrTheory}.
At present, the strongest constraint on the $\LL$ interaction
is provided by the $\LL$ bond energy of the \DLH,
$\HeLL{6}$, observed in the Nagara event~\cite{Nagara}.
The $\LL$ bond energy represents the strength of the $\LL$ interaction,
and is defined as
$\Delta B_{\LL}\equiv
S_{\LL}(^{\;\;A}_{\LL}Z)
-2S_{\Lambda}(^{A-1}_{\,\;\;\;\;\Lambda}{Z})$,
where $S_{\LL}$ and $S_\Lambda$ are the separation 
(binding) energies of $\LL$ and $\Lambda$.
The bond energy of $\HeLL{6}$ is found to be
$\Delta B_{\LL}(\HeLL{6}) =0.67 \pm 0.12~\MeV$~\cite{E373}.
This bond energy can be fitted by a $\LL$ interaction
with the low-energy scattering parameters of
$(a_0,r_\mathrm{eff})=(-0.44~\mathrm{fm}, 10.1~\mathrm{fm})$~\cite{Hiyama2010},
where $a_0$ and $r_\mathrm{eff}$ are
the scattering length and the effective range, respectively.

The $\LL$ interaction plays a crucial role also in neutron star physics.
With most of the well-known attractive $\Lambda N$ two-body interactions,
$\Lambda$ hyperon mixing is calculated to take place
in neutron star matter at $(2-4)\rho_0$ and the equation of state (EOS)
is softened~\cite{HypEOS}. Consequently it is hard 
to support massive neutron stars
with masses around two solar mass~\cite{MassiveNS}.
In order to solve this problem, known as the hyperon puzzle,
several mechanisms have been proposed so far.
One of the natural ways is to introduce repulsive three-baryon
interactions~\cite{3BR,Tsubakihara2013}.
For example, it is possible to support massive neutron stars
by introducing repulsive three-body contact couplings
in RMF models~\cite{Tsubakihara2013}.
It should be noted that the three-baryon repulsion needs to operate
also among $\LL{N}$ and $\LL{\Lambda}$.
Otherwise $\Lambda$ matter becomes more stable
than nuclear matter with hyperon mixing at high densities.
The $\LL{N}$ three-baryon interaction will cause
effective density-dependent $\LL$ and $\Lambda{N}$ interactions
in nuclear matter.
If we can observe and uniquely identify many \DLH\ in a wide mass region,
it would be possible to deduce the density dependence of the $\LL$ interaction
and the underlying $\LL{N}$ three-baryon interaction.

The most efficient reaction to form \DLH\ is the $\Xi^-$ absorption at rest
in nuclei.
\DLH\ formation proceeds in the following steps.
First $\Xi^-$ particles are produced in the $(K^-,K^+)$ reactions
on nuclei or protons~\cite{Danysz1963,E176,Nagara,E373,E07}
or $\bar{p}$-nucleus collisions~\cite{PANDA}.
The produced $\Xi^-$ particle is absorbed in a nucleus,
and converted to two $\Lambda$ particles via the 
$\Xi^-p \to \LL$ reaction in the nucleus.
If two $\Lambda$s are trapped in the nucleus in the preequilibrium stage,
a double $\Lambda$ compound nucleus (\DLC) is formed~\cite{DLC}.
The compound nucleus deexcite by emitting
nucleons, $\Lambda$s, $\alpha$s and other clusters.
When these two $\Lambda$ particles occasionally stay in the same fragment
with an excitation energy below the particle emission threshold,
a \DLH\ is formed.
When the sequential weak decay of the \DLH\ 
is observed, one can identify that a $S=-2$ nucleus is formed.
It is also expected to produce and detect \DLH\ in
heavy-ion collisions~\cite{HIC},
while there is no clear evidence of \DLH\ formation in these reactions yet.

\begin{table}[htbp]
\caption{Uniquely identified double $\Lambda$ hypernuclear formation
events~\cite{Danysz1963,E176,Nagara,E373}.
We also show the possible formation channels in the Mino event,
a new event observed in the J-PARC E07 experiment~\cite{E07}.
}\label{Tab:DLH}
\begin{center}
\begin{tabular}{c|l|c|c}
\hline
\hline
Experiment & Reaction & $\dBLL~(\MeV)$ & Ref.\\
\hline
CERN	& 
$\Xi^-+\nuc{12}{C} \to \nucLL{\;10}{Be}+d+n$
& $1.3 \pm 0.4$
& \cite{Danysz1963}\\
\hline
KEK-E176 &
$\Xi^-+\nuc{14}{N} \to \nucLL{\;13}{B}+p+n$
& $0.6 \pm 0.8$
&\cite{E176}\\
\hline
Nagara &
$\Xi^-+\nuc{12}{C} \to \nucLL{\,\;\;6}{He}+\alpha+t$
& $0.67 \pm 0.17$
&\cite{Nagara,E373}\\
\hline
Demachi-Yanagi &
$\Xi^-+\nuc{12}{C} \to \nucLL{\;10}{Be}^*+d+n$
& $-1.52 \pm 0.15$
& \cite{E373}\\
\hline
\hline
\multirow{3}{*}{Mino}
&$\Xi^-+\nuc{16}{O} \to \nucLL{\;10}{Be}+\alpha+t$
&$1.63\pm 0.14$ 
&\multirow{3}{*}{\cite{E07}} \\
&$\Xi^-+\nuc{16}{O} \to \nucLL{\;11}{Be}+\alpha+d$
&$1.87\pm 0.37$\\
&$\Xi^-+\nuc{16}{O} \to \nucLL{\;12}{Be}^*+\alpha+p$
&$-2.7\pm 1.0$\\
\hline
\hline
\end{tabular}
\end{center}
\end{table}

Until now, several experiments have been performed to find \DLH,
and four of them have been uniquely identified from $\Xi^-$ absorption in nuclei
as summarized in Table~\ref{Tab:DLH}~\cite{Danysz1963,E176,Nagara,E373};
$\nucLL{\;10}{Be}$ from $\Xi^-+\nuc{12}{C}$ reaction at CERN~\cite{Danysz1963},
$\nucLL{\;13}{B}$ from $\Xi^-+\nuc{14}{N}$ in KEK-E176 experiment~\cite{E176},
$\Lambpha$ from $\Xi^-+\nuc{12}{C}$ (Nagara event)~\cite{Nagara}
and 
$\nucLL{\;10}{Be}^*$ from $\Xi^-+\nuc{12}{C}$ (Demachi-Yanagi event)~\cite{E373}
in KEK-E373 experiment.
Another report on $\Lambpha$~\cite{Prowse1966}
was questioned~\cite{Dalitz1989}
and found to be not consistent with the Nagara event~\cite{Nagara,E373}.
It should be noted that the identifications of the above four \DLH\ 
rely on the consistency between the events.
For example, $B_{\LL}$ values of $\nucLL{\;10}{Be}$
in Refs.~\cite{Danysz1963} and ~\cite{E373}
can be consistent by assuming
the channel of $\nucLL{\;10}{Be}\to \nucL{9}{Be}^*+p+\pi^-$
in the weak decay in the event of Ref.~\cite{Danysz1963}
and the formation of the excited state $\nucLL{\;10}{Be}^*$
in the Demachi-Yanagi event~\cite{E373}.
In order to further observe \DLH, 
the J-PARC-E07 experiment has been carried out.
While the analysis is still on-going,
a new \DLH\ formation event is discovered recently,
$\nucLL{}{Be}$ from $\Xi^-+\nuc{16}{O}$ (Mino event)~\cite{E07}.
We summarize the uniquely identified \DLH\ events
in Table~\ref{Tab:DLH}. We also show the candidate fragmentation
reactions in the Mino event.

Since further events are expected to be observed from the J-PARC-E07
and future experiments,
it would be possible to perform statistical analysis of
fragment formation events from $\Xi^-$ absorption at rest in light nuclei.
Statistical decay of \DLC\ was studied
by using a canonical fragmentation model~\cite{Yamamoto1994},
a sequential binary statistical decay model~\cite{Hirata1999},
and a microcanonical fragmentation model~\cite{Lorente2011}.
We also note that the $\Xi^-$ absorption reaction in $\nuc{12}{C}$
was analyzed by using the direct reaction model in Ref.~\cite{Yamada1997}.
In Ref.~\cite{Hirata1999}, the $\Xi^-$ absorption reaction in $\nuc{12}{C}$
was analyzed in a combined framework of a transport model
and a statistical decay model of hypernuclei.
The formation probability of the \DLC\ ($\nucLL{\;13}{B}^*$)
was evaluated to be around $\PDLC \simeq 30~\%$
in the preequilibrium stage by using the antisymmetrized molecular dynamics
(AMD) transport model calculation~\cite{Hirata1999},
the sum of branching ratios to form \DLH\ from \DLC\ 
in the statistical decay was found to be around $\PbrTot = 60~\%$,
and then the total \DLH\ formation
probability was found to be around 
$\PDLHtot = \PDLC \times \PbrTot \simeq 18~\%$
after the statistical decay.
This analysis was performed before the discovery
of the Nagara event~\cite{Nagara},
and a strongly attractive $\LL$ interaction was adopted,
$\dBLL(\nucLL{\;13}{B})=4.9~\MeV$,
as suggested by the KEK-E176 experiment~\cite{E176}.
After the Nagara event, the KEK-E176 event was reinterpreted
and the $\LL$ bond energy is now considered to be
$\dBLL(\nucLL{\;13}{B})=0.6 \pm 0.8~\MeV$.
With these updated less attractive $\LL$ interaction,
it would be possible to predict 
the \DLH\ formation probabilities in a more reliable manner.

It should be noted that the formation probability of \DLC ($\PDLC$)
depends on the definition of compound nucleus formation,
and in practice
it depends on the transport model adopted in describing the dynamical stage.
For example, the \DLC\ formation probability
in stopped $\Xi^-$ absorption in $\nuc{12}{C}$
is calculated to be smaller, $\PDLC(\mathrm{AMD{\hyph}QL})=16~\%$,
in AMD with additional quantum fluctuations (AMD-QL)~\cite{Hirata1999}.
In addition to the difference in $\PDLC$,
emission of nucleons and light clusters in the dynamical stage
would modify the mass dependence of the formation probability of \DLC.
In Subsec.~\ref{Sec:C12Xi}, we compare the statistical decay model results
from \DLC\ ($\nucLL{\;13}{B}^*$)
and AMD-QL results with the statistical decay in Ref.~\cite{Hirata1999}.

In this article, we discuss the formation of \DLH\ 
from the statistical decay of \DLC\ 
formed via the $\Xi^-$ absorption at rest in nuclei.
Specifically, we concentrate on the target nuclei
$\nuc{12}{C}$, $\nuc{14}{N}$ and $\nuc{16}{O}$.
These are the main light component of emulsion,
and some \DLH\ have been reported to be formed.
We mainly discuss the formation probabilities 
$\Lambpha$ and $\nucLL{\;10}{Be}$ from $\Xi^-+\nuc{12}{C}$,
$\nucLL{\;13}{B}$ from $\Xi^-+\nuc{14}{N}$,
and 
$\nucLL{\;11}{Be}$ from $\Xi^-+\nuc{16}{O}$.
We also discuss
the dependence of the formation probabilities on the $\LL$ bond energy,
and the decay of $\Xi$ hypernucleus formed
in the $\nuc{7}{Li}(K^-,K^+)$ reaction.

This paper is organized as follows.
In Sec.~\ref{Sec:SDM}, we provide a general idea of statistical decay
of compound hypernuclei.
In Sec.~\ref{Sec:results}, we evaluate \DLH\ formation probabilities
from $\Xi$ absorption reaction at rest in $^{12}$C, $^{14}$N and $^{16}$O.
In Sec.~\ref{Sec:He6Xi}, we discuss $\HLL{5}$ formation
probability from $\Xi^--^6$He bound state
at various values of the $\Xi^-$ binding energy.
Summary and discussion are given in Sec.~\ref{Sec:Summary}.


\section{Statistical decay of double $\Lambda$ compound nuclei
and hypernuclear binding energies}
\label{Sec:SDM}

In this paper, the statistical decay of 
\DLC\ is calculated by using a sequential binary statistical decay model
(SDM)~\cite{SDM}.
In SDM, an excited nucleus $1$ is assumed to decay to nuclei $2$ and $3$
with the decay rate of
\begin{align}
\Gamma_{1\to23} dE_2 dE_3 = \frac{\rho_2(E_2,J_2)\rho_3(E_3,J_3)}{2\pi\rho_1(E_1,J_1)}\sum_{L,J_{23}} T_L dE_2 dE_3 ,
\label{Eq:SDM}
\end{align}
where $E_i$ and $J_i$ denote the excitation energy and the angular momentum
of the $i$-th nucleus.
We adopt the back-shifted Fermi gas model
to evaluate the level density $\rho_i(E_i,J_i)$~\cite{SDM,FaiRandrup1982}.
Since we are interested in decays of compound nuclei in equilibrium,
density of narrow excited levels needs to be considered
and level density is reduced at excitation energies
above the threshold for charge neutral particle emission
and above the Coulomb barrier
for charged particle emission~\cite{FaiRandrup1982}.
The factor $T_L=\Theta(L_c-L)$ is the transmission coefficient
of the partial wave $L$ in the fusion reaction $2+3\to1$
with the incident energy corresponding to the excitation energies
and the $Q$-value.
We assume the strong absorption in the inverse fusion process $2+3\to1$,
and take the form $T_L=\Theta(L_c-L)$ with $L_c$ being
the maximum orbital angular momentum of the fusion.
Statistical decays are assumed to be binary and to proceed
until all the fragments are in their ground states.
When combined with the transport models describing the preequilibrium stage,
the SDM has been found to work well
for intermediate energy heavy-ion collisions~\cite{MD_SDM}
and for hypernuclear formation reactions~\cite{Hirata1999,Nara1995}.
We would also like to mention here that the combined framework
of the transport model and SDM is
successfully applied to the light-ion induced reaction,
$p+\nuc{12}{C}$ at 45 MeV,
where the excitation energy of the compound nucleus is similar
to the $\Xi^-$ absorption reaction at rest in nuclei
and essentially the same SDM program is used~\cite{Hirata1999}.

\begin{figure}[bthp]
\begin{center}
\includegraphics[width=0.50\textwidth]{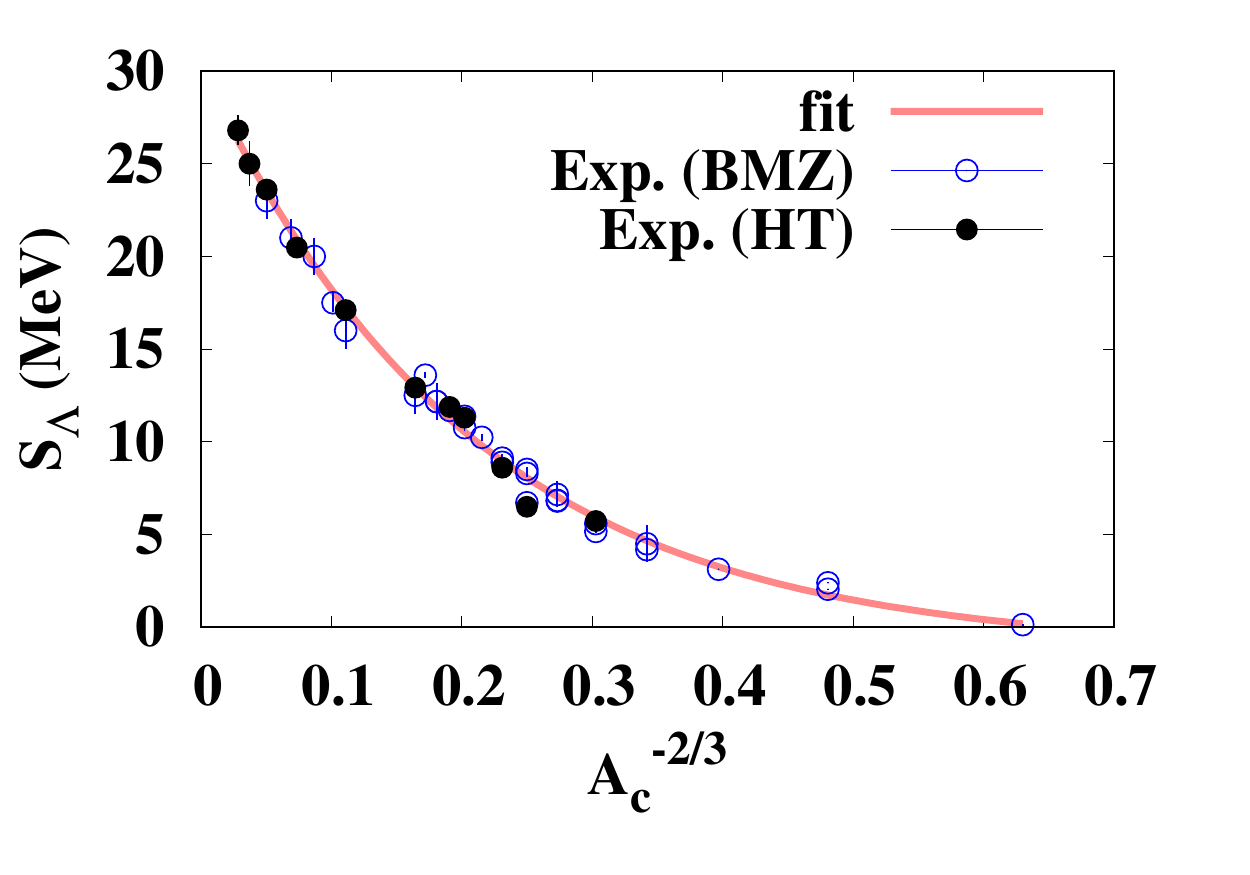}%
\includegraphics[width=0.50\textwidth]{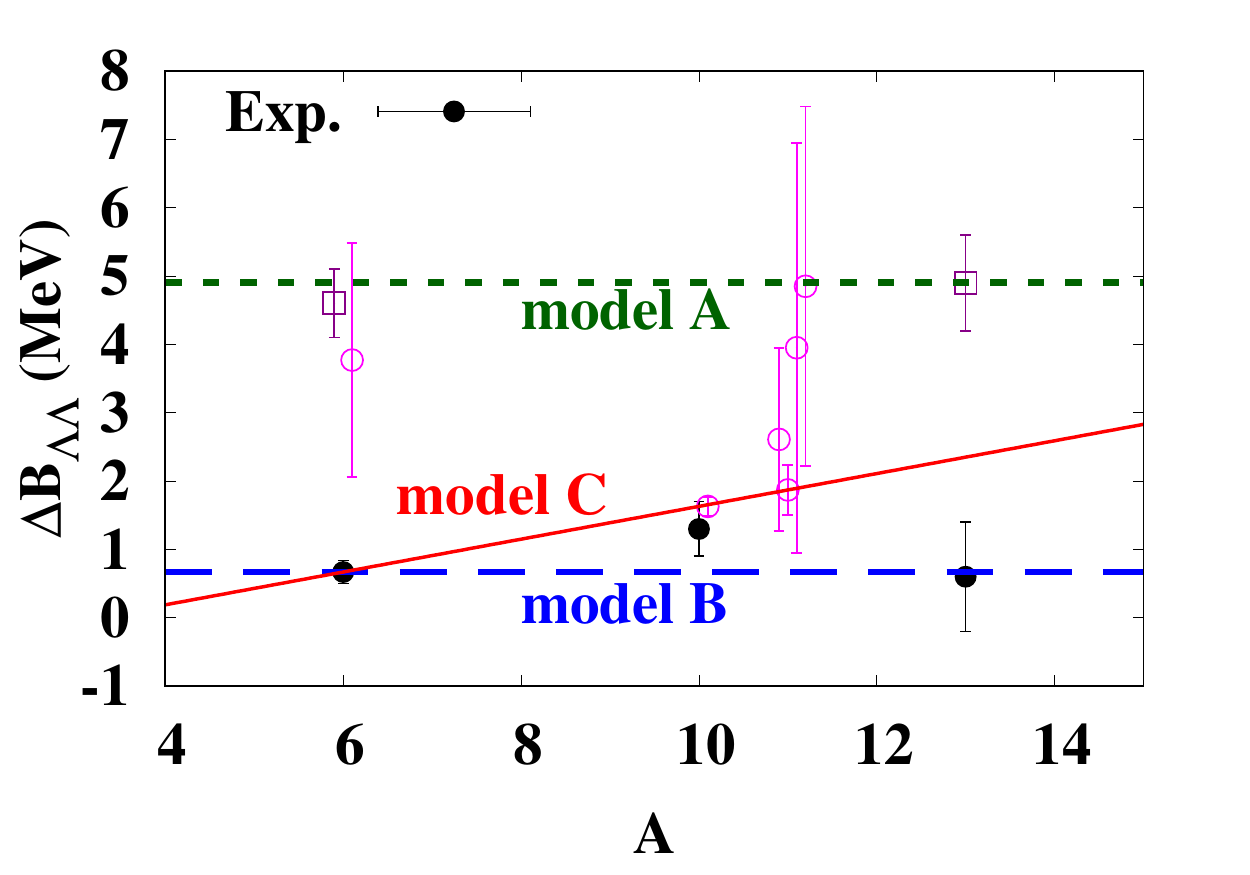}%
\end{center}
\caption{
Left: Separation energy of single hypernuclei.
Data summarized in \cite{BMZ} (open circles) and \cite{HT} (filled circles)
are shown by symbols,
and the results of fit is shown by the solid curve.
Right: $\LL$ bond energy of \DLH.
Filled circles show the results of uniquely identified ground state
$\LL$ bond energy $\dBLL$.
Dotted, dashed and solid lines show $\dBLL$ in the models A, B and C,
respectively.
Open circles and squares show experimental data of
$\dBLL$ without unique identifications and
rejected values, respectively.
}
\label{Fig:SL}
\end{figure}

The essential inputs of SDM is the binding energies of nuclei,
which would be formed during the sequential decay processes.
While the binding energies of normal nuclei are well known,
information on single and double hypernuclear binding energies is limited.
Then we have constructed the mass table based on the following assumptions.
First, all existing normal nuclei plus $\Lambda$ and $\LL$
are assumed to form single and double hypernuclei, respectively.
Normal nuclei include those whose ground states are resonance
and unstable to particle emission.
We also consider dineutron ($nn$) and $\nuc{2}{He}$ ($pp$) as resonance nuclei,
and their energies are assumed to be $100~\mathrm{keV}$ above the threshold,
in order to mimic three-body decays.
Second, for single hypernuclei,
we adopt the separation energy if measured.
We use the separation energy data summarized
by Bando et al. (BMZ)~\cite{BMZ}
and by Hashimoto and Tamura (HT)~\cite{HT}.
We add $0.5~\MeV$ to the separation energies measured in $(\pi^+,K^+)$
experiments~\cite{HT} to take account of the recalibration
as shown in Ref.~\cite{Gogami2016}.
If not measured,
we adopt the $\Lambda$ separation energy $S_\Lambda$,
parameterized as a function of the mass number
and fitted to the observed separation energies,
\begin{align}
S_\Lambda=\frac{c_0 (A-1)}{A}\frac{1-c_1 x}{1+c_2x+c_3x^2} ,
\end{align}
where $A$ denotes the hypernuclear mass number and $x=1/(A-1)^{2/3}$.
By fitting to the BMZ and HT data, we determined the parameters of
$c_0=30.18~\MeV$, $c_1=1.52$, $c_2=3.17$ and $c_3=4.38$.
Results of the fit is shown in Fig.~\ref{Fig:SL}.
Thirdly, for \DLH, 
the $\LL$ bond energy is given as a linear function of the mass number.
We compare the results of the three models,
$\Delta B_{\Lambda \Lambda}=4.9~\MeV$ (model A),
$\Delta B_{\Lambda \Lambda}=0.67~\MeV$ (model B),
and 
$\Delta B_{\Lambda \Lambda}(A=6)=0.67~\MeV$
and
$\Delta B_{\Lambda \Lambda}(A=11)=1.87~\MeV$ (model C).
These model parameters are determined by
the old rejected value of $\dBLL$ from E176 experiment~\cite{E176a} 
(model A),
the Nagara event (model B),
and by the Nagara and Mino events (model C).

\section{Double hypernuclear formation from $\Xi^-$ absorption in nuclei}
\label{Sec:results}

We now discuss \DLH\ formation from $\Xi^-$ absorption reaction at rest
in $\nuc{12}{C}$, $\nuc{14}{N}$ and $\nuc{16}{O}$.
We assume that $\Xi^-$ is absorbed from the $3D$ atomic orbit,
so the $\Xi^-$ binding energy is
$B=\mu\alpha^2Z^2/2n^2=0.126, 0.174$ and $0.230~\MeV$
for $\nuc{12}{C}$, $\nuc{14}{N}$ and $\nuc{16}{O}$ targets, respectively,
where $\mu$ is the reduced mass, $Z$ is the atomic number of the target,
and $n=3$ for the $3D$ orbit.
A part of the released energy in the conversion process,
$M_{\Xi^-}+M_p-2M_\Lambda=28.62~\MeV$,
is used by the separation energy of the proton,
$15.96, 7.55$ and $12.13~\MeV$,
and thus the energies from the threshold,
$\nuc{11}{B}+\LL$, $\nuc{13}{C}+\LL$ and $\nuc{15}{N}+\LL$,
are obtained as 
$12.66, 21.07$ and $16.49~\MeV$,
for $\nuc{12}{C}$, $\nuc{14}{N}$ and $\nuc{16}{O}$ targets, respectively.
These energies are summarized in Table~\ref{Tab:DLC}.

\begin{table}[htbp]
\caption{Relevant energies in stopped $\Xi^-$ absorption reaction at rest in 
$\nuc{12}{C}$, $\nuc{14}{N}$ and $\nuc{16}{O}$.
We show the target binding energies ($B_T$),
the proton separation energies ($S_p$)
and the energy from the $\LL$ emission threshold ($E_\mathrm{th}$).
We also show the case of the $\nuc{6}{He}$ target, a core nucleus of
$\nucX{7}{H}$, which may be formed in the $\nuc{7}{Li}(K^-,K^+)$ reaction. 
}\label{Tab:DLC}
\begin{center}
\begin{tabular}{c|r|c|c|c}
\hline
\hline
Target
& $B_T (\MeV)$
& $S_p~(\MeV)$
& $E_\mathrm{th}~(\MeV)$
& $B_\Xi~(\MeV)$\\
\hline
$\nuc{12}{C}$ & 92.165	& 15.957 & 12.569 - $B_\Xi$ & 0.126 \\
$\nuc{14}{N}$ & 76.047	&  7.551 & 21.065 - $B_\Xi$ & 0.174 \\
$\nuc{16}{O}$ & 127.624	& 12.127 & 16.489 - $B_\Xi$ & 0.230 \\
\hline
$\nuc{6}{He}$ & 29.268	& 23.482 & 5.137 - $B_\Xi$  & 0-4	\\
\hline
\hline
\end{tabular}
\end{center}
\end{table}

\begin{figure}[htbp]
\begin{center}
\includegraphics[width=0.5\textwidth]{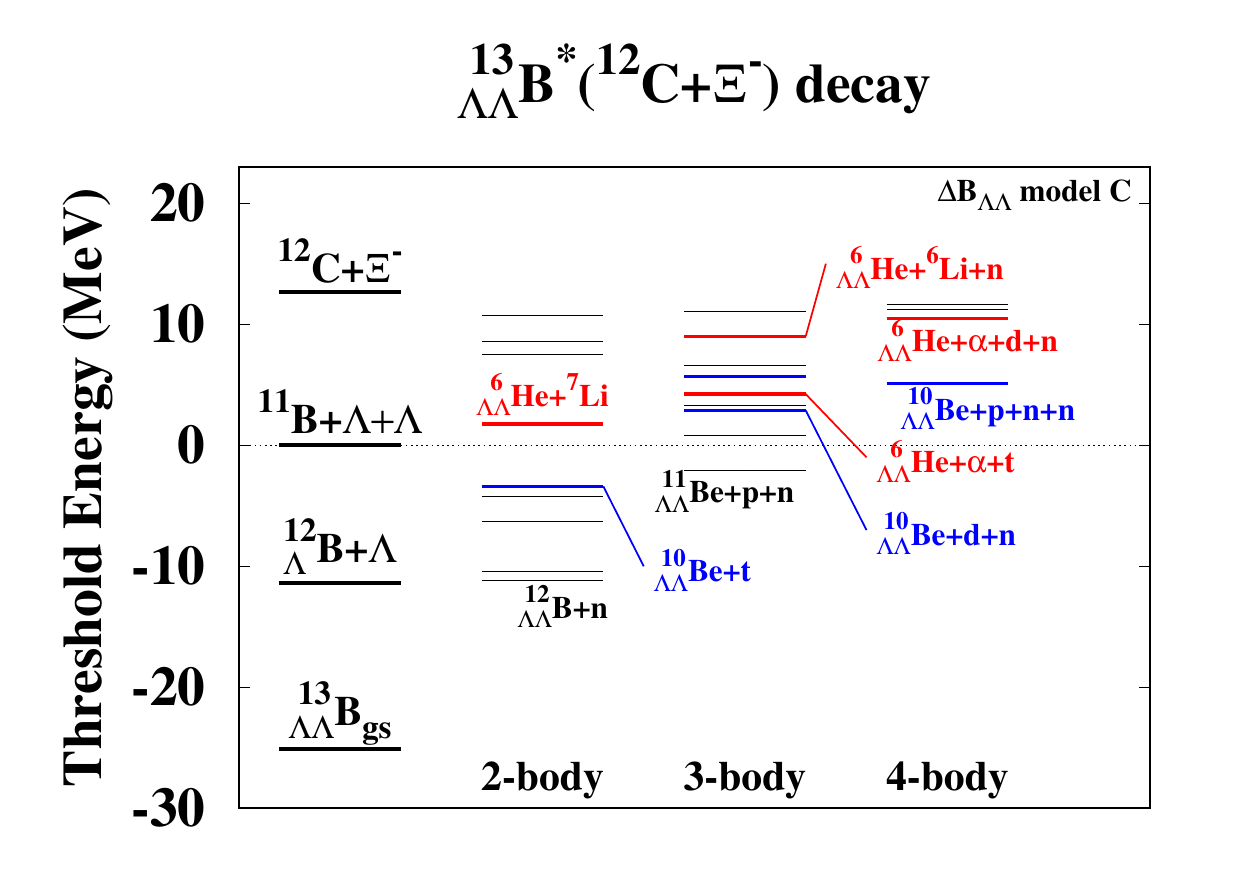}%
\includegraphics[width=0.5\textwidth]{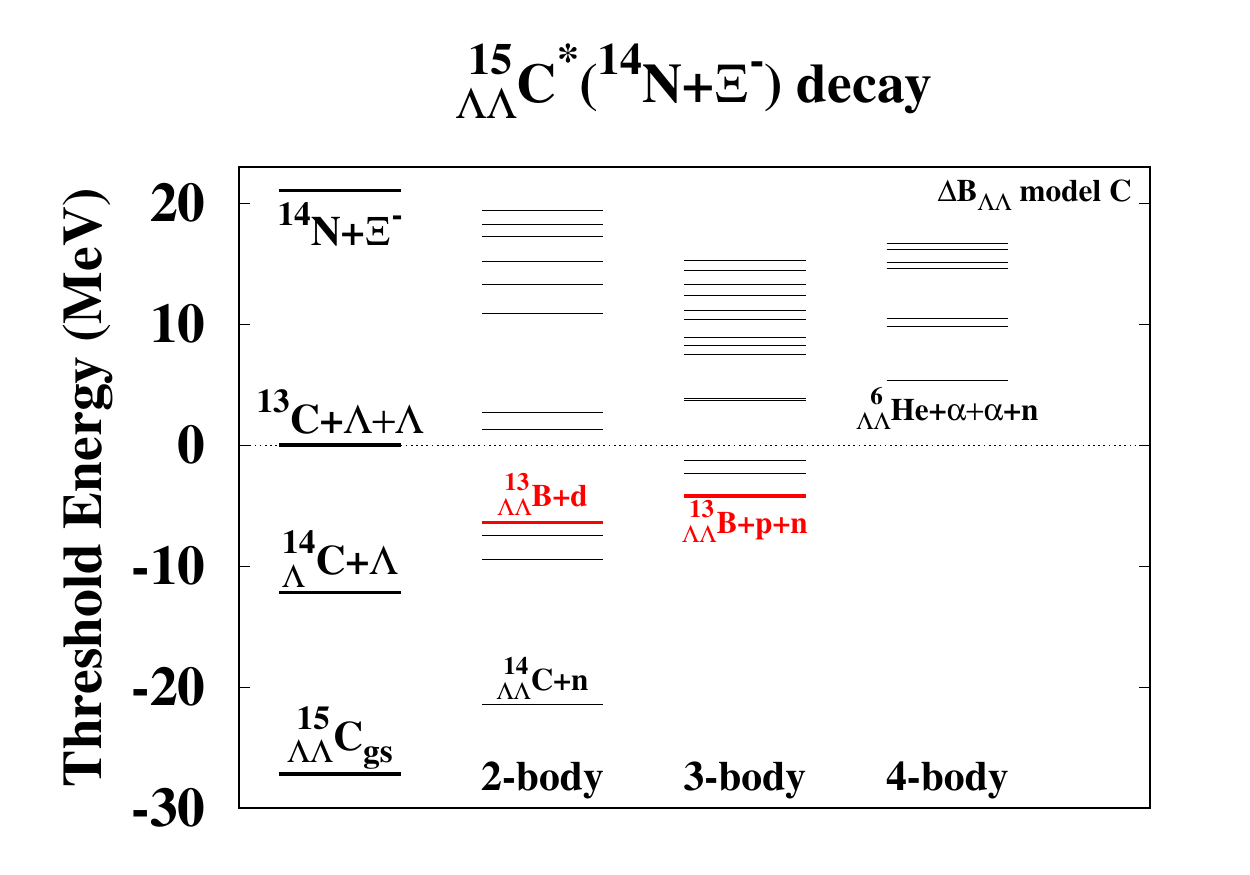}%
\\
\includegraphics[width=0.5\textwidth]{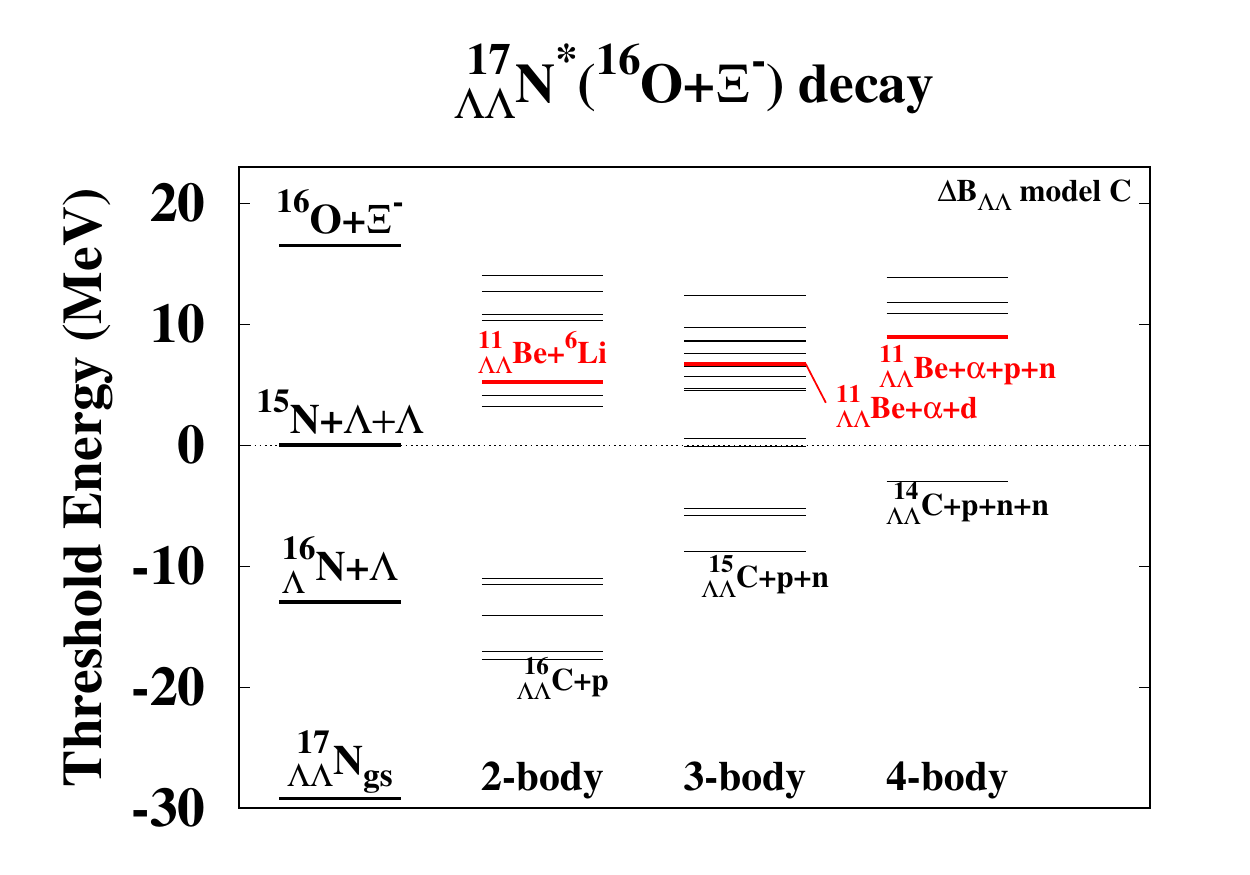}%
\end{center}
\caption{
Open channels including \DLH\ from $\Xi^-$ absorption
in $\nuc{12}{C}$, $\nuc{14}{N}$ and $\nuc{16}{O}$.
}
\label{Fig:thrCNO}
\end{figure}

Open channels including \DLH\ from the $\Xi^-$ absorption reactions are shown 
in Fig.~\ref{Fig:thrCNO}.
Many channels including 2-, 3- and 4-body
can be populated in the final state.
We have performed the statistical decay model calculation
from the \DLC\ until all the fragments are deexcited to the ground state.
In order to handle variety of decay path,
the decay channel, excitation energies and angular momenta of 
daughter fragments are chosen in the Monte-Carlo method.

\begin{table}[htbp]
\caption{Formation probabilities of \DLH\ in the statistical decay model
from the \DLC,
$\nucLL{\;13}{B}^* (\Xi^-+\nuc{12}{C})$,
$\nucLL{\;15}{C}^* (\Xi^-+\nuc{14}{N})$,
and 
$\nucLL{\;17}{N}^* (\Xi^-+\nuc{16}{O})$.
The probabilities ($\Pbr$) are given in \%.
The "-" symbols show that the \DLH\ are either unbound,
energetically not accessible, or not populated in the Monte-Carlo sampling
of the decay paths in the SDM calculation.}
\label{Tab:Adist}
\begin{center}
\begin{tabular}{r|rrr|rrr|rrr}
\hline
\hline
$^{\;\;A}_{\LL}Z$ & \multicolumn{3}{c}{$\nuc{12}{C}$}
					& \multicolumn{3}{|c}{$\nuc{14}{N}$}
								& \multicolumn{3}{|c}{$\nuc{16}{O}$} \\
\hline
model		& A	& B	& C	& A	& B	& C	& A	& B	& C	\\
\hline
sum			& 71.2	& 41.0	& 50.6	& 56.2	& 24.6	& 37.3	& 79.1	& 59.2	& 71.5	\\
\hline
$\HLL{4}$		&0.03	&-	&-	&0.01	&0.002	&0.001	&0.003	&-	&-      \\
$\HLL{4}$		&0.6	&0.4	&0.4	&0.02	&0.006	&0.005	&0.004	&0.006	&-      \\
\hline
$\Lambpha$		&4.3	&1.6	&1.3	&3.6	&1.1	&1.1	&0.1	&0.07	&0.05   \\
$\HeLL{7}$		&1.6	&0.4	&0.4	&0.1	&0.1	&0.1	&0.02	&-	&-      \\
$\HeLL{8}$		&0.6	&0.3	&0.2	&0.01	&0.004	&0.002	&0.002	&-	&-      \\
\hline
$\nucLL{\,\;\;7}{Li}$		&0.4	&-	&-	&0.01	&-	&-	&0.008	&-	&-      \\
$\nucLL{\,\;\;8}{Li}$		&3.9	&1.6	&1.7	&0.07	&0.03	&0.02	&0.02	&-	&-      \\
$\nucLL{\,\;\;9}{Li}$		&0.6	&0.7	&0.7	&0.5	&0.2	&0.2	&0.2	&0.1	&0.1    \\
$\nucLL{\;10}{Li}$	&0.2	&0.01	&0.05	&0.1	&0.08	&0.07	&-	&-	&-      \\
\hline
$\nucLL{\,\;\;9}{Be}$		&-	&-	&-	&0.2	&0.02	&0.02	&-	&-	&-      \\
$\nucLL{\;10}{Be}$	&18.7	&5.5	&8.0	&2.6	&1.4	&2.0	&0.4	&0.08	&0.09   \\
$\nucLL{\;11}{Be}$	&27.1	&18.9	&24.3	&1.2	&0.5	&0.6	&1.1	&0.2	&0.3    \\
$\nucLL{\;12}{Be}$	&0.8	&1.4	&1.4	&1.1	&0.4	&0.5	&0.6	&0.3	&0.4    \\
$\nucLL{\;13}{Be}$	&-	&-	&-	&0.03	&0.05	&0.04	&-	&-	&-      \\
\hline
$\nucLL{\;11}{B}$		&8.7	&4.3	&6.3	&0.3	&0.1	&0.1	&0.2	&0.003	&0.02   \\
$\nucLL{\;12}{B}$		&3.6	&6.0	&5.9	&16.4	&3.4	&8.5	&3.3	&1.9	&2.2    \\
$\nucLL{\;13}{B}$		&-	&-	&-	&18.5	&11.2	&15.5	&2.0	&0.7	&1.0    \\
$\nucLL{\;14}{B}$		&-	&-	&-	&0.02	&0.04	&0.05	&0.1	&0.04	&0.07   \\
$\nucLL{\;15}{B}$		&-	&-	&-	&-	&-	&-	&0.1	&0.03	&0.09   \\
\hline
$\nucLL{\;12}{C}$		&-	&-	&-	&0.1	&0.02	&0.04	&-	&-	&-      \\
$\nucLL{\;13}{C}$		&-	&-	&-	&10.0	&5.4	&7.5	&0.8	&0.09	&0.2    \\
$\nucLL{\;14}{C}$		&-	&-	&-	&1.3	&0.7	&1.0	&59.3	&38.5	&52.8   \\
$\nucLL{\;15}{C}$		&-	&-	&-	&-	&-	&-	&8.0	&12.4	&10.0   \\
$\nucLL{\;16}{C}$		&-	&-	&-	&-	&-	&-	&0.1	&0.3	&0.2    \\
\hline
$\nucLL{\;15}{N}$		&-	&-	&-	&-	&-	&-	&2.5	&3.6	&3.4    \\
$\nucLL{\;16}{N}$		&-	&-	&-	&-	&-	&-	&0.4	&0.9	&0.6    \\
\hline
\hline
\end{tabular}
\end{center}
\end{table}

In Table~\ref{Tab:Adist}, we summarize 
the formation probabilities of \DLH\ in the SDM calculation 
from the \DLC,
$\nucLL{\;13}{B}^* (\Xi^-+\nuc{12}{C})$,
$\nucLL{\;15}{C}^* (\Xi^-+\nuc{14}{N})$,
and 
$\nucLL{\;17}{N}^* (\Xi^-+\nuc{16}{O})$.
The total \DLH\ formation probability
in SDM is calculated to be $\PbrTot = (25-80)~\%$.
If we assume that the \DLC\ formation probability
is common in the three targets and set to be
$\PDLC = 30~\%$~\cite{Hirata1999},
the total \DLH\ formation probability is
$\PDLHtot = \PDLC \times \PbrTot = (7.5-24)~\%$,
which is larger than the lower limit of the $2\Lambda$ trapping probability
(double and twin hypernuclear formation), $4.8~\%$ for light nuclei,
evaluated by the KEK-E176 collaboration~\cite{E176a,E176b}.
By comparison,
the calculated \DLH\ formation probability is larger than
the $2\Lambda$ trapping probability captured by light nuclei,
$5.0 \pm 1.7~\%$~\cite{E373trap}.
The small $2\Lambda$ trapping probability data
may suggest small bond energies as in the Model B.

\subsection{$\Xi^-$ absorption in $^{12}\mathrm{C}$}
\label{Sec:C12Xi}

\begin{figure}[htbp]
\begin{center}
\includegraphics[width=0.7\textwidth]{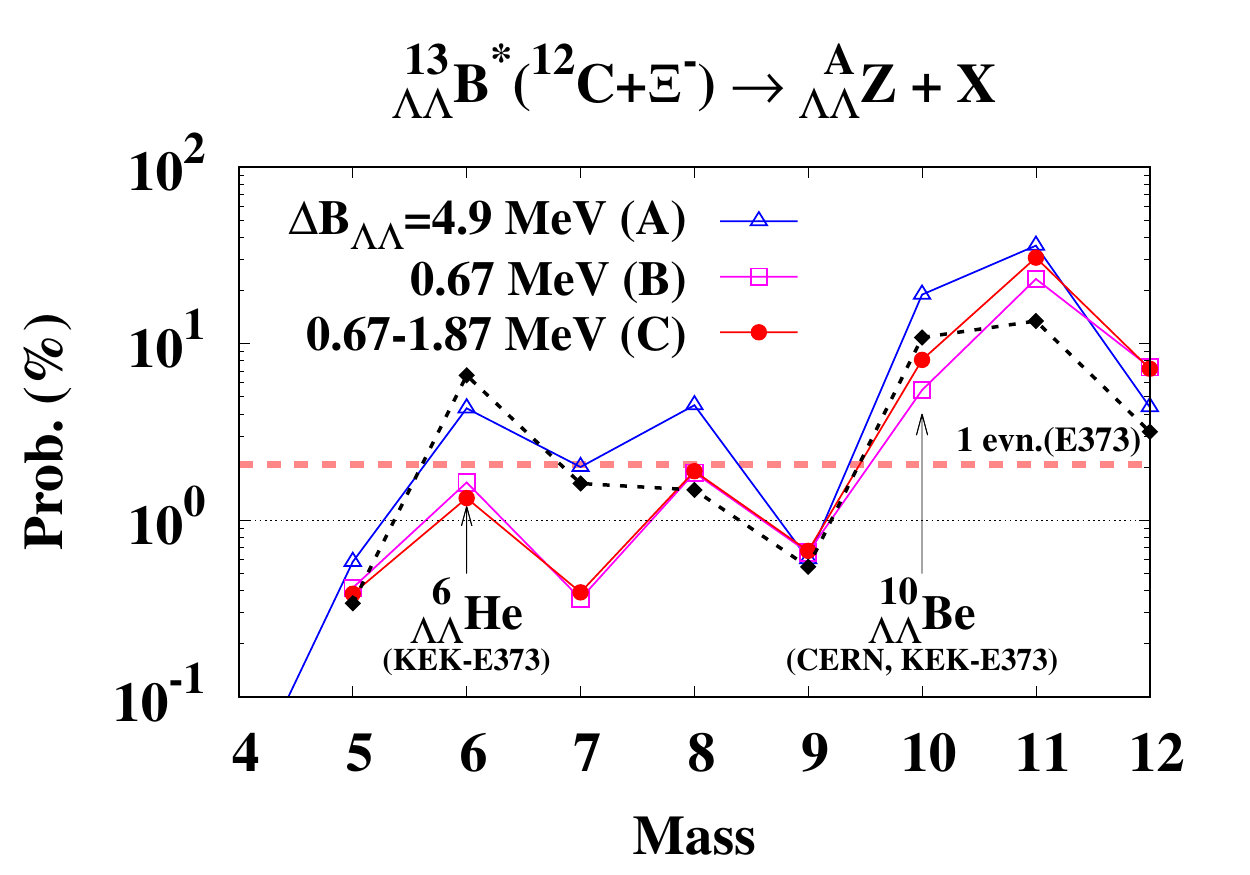}%
\end{center}
\caption{
\DLH\ mass distribution ($\Pbr$)
obtained in the statistical decay model starting from
the \DLC\ ($\nucLL{\;13}{B}^*$)
assumed to be formed from the $\Xi^-$ absorption at rest
from the $3D$ atomic orbit in $\nuc{12}{C}$.
Results of the model A (triangles), B (squares) and C (circles)
for the $\LL$ bond energy $\dBLL$ are compared.
Diamonds show the results of AMD-QL with strong fluctuations
on \DLH\ formation probability~\cite{Hirata1999} divided by 0.3.}
\label{Fig:C12Xi}
\end{figure}

First we discuss the $\Xi^-$ absorption in $\nuc{12}{C}$,
where $\HeLL{6}$~\cite{Nagara} and $\nucLL{\;10}{Be}$~\cite{Danysz1963,E373}
were observed.
In Fig.~\ref{Fig:C12Xi}, we show the \DLH\ mass distribution
obtained in the SDM calculation of the \DLC, $\nucLL{\;13}{B}^*$,
assumed to be formed in the $\Xi^-$ absorption in $\nuc{12}{C}$.
We find that \DLH\ with $A=11$ is most frequently formed;
$\Pbr=(18.9-27.1)~\%$ for $\nucLL{\;11}{Be}$ and
$\Pbr=(4.3-8.7)~\%$ for $\nucLL{\;11}{B}$.
Next frequently formed \DLH\ is $\nucLL{\;10}{Be}$
with the probability of $\Pbr=(5.5-18.7)~\%$.
While the $\dBLL$ model dependence is not significant for heavier ($A\geq 11$)
hypernuclei, small $\dBLL$ in models B and C suppress formation probabilities
of lighter hypernuclei.
For example, the formation probability of $\HeLL{6}$
is calculated to be $\Pbr=4.3~\%$ in model A,
while it becomes $\Pbr=1.6~\%$ and $1.3~\%$ in models B and C, respectively.


\begin{figure}[htbp]
\begin{center}
\includegraphics[width=0.7\textwidth]{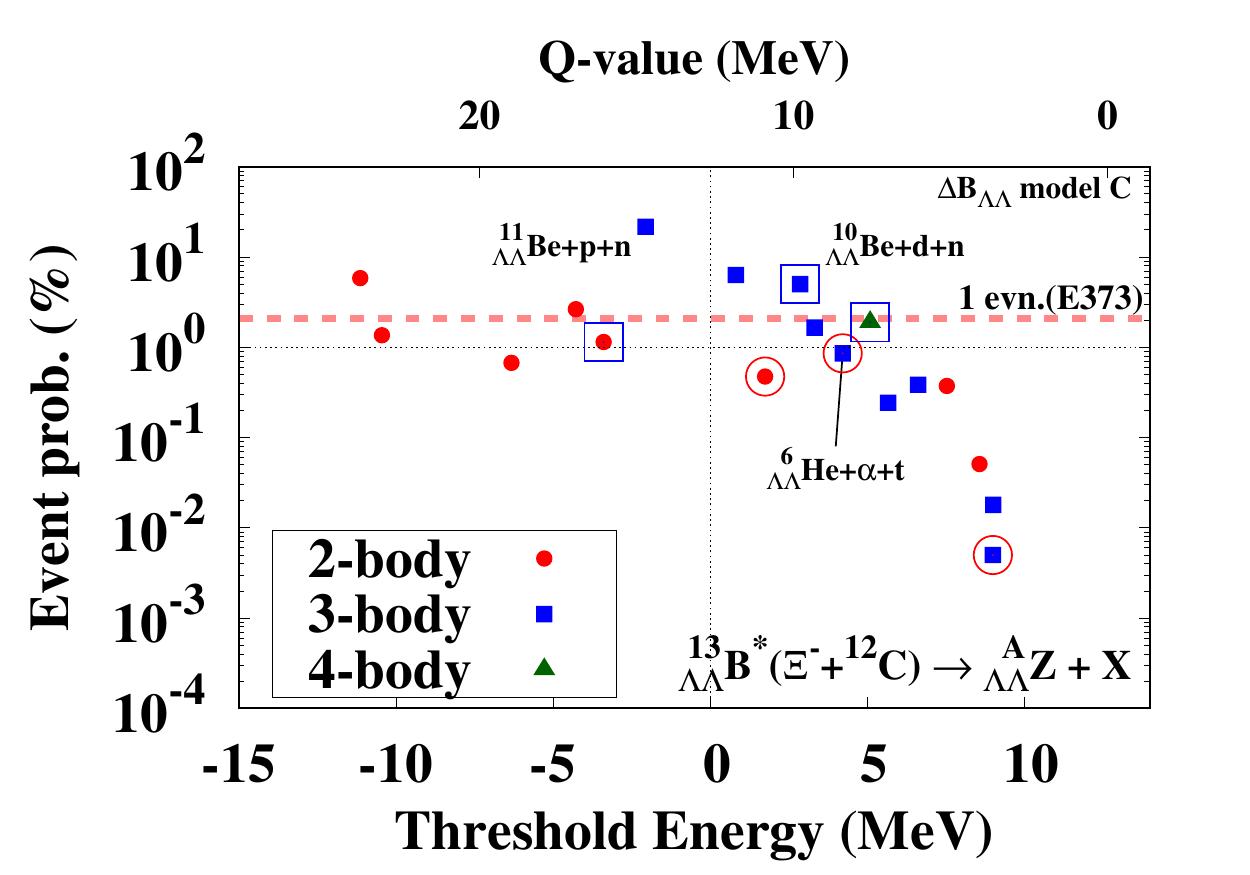}%
\end{center}
\caption{
Threshold energy dependence of the event probabilities ($\Pbr$)
obtained in SDM from $\nucLL{\;13}{B}^*$ ($\Xi^-+\nuc{12}{C}$).
Filled circles, squares and triangle show the event probability in SDM
as a function of the threshold energy
in 2-body, 3-body and 4-body channels, respectively.
Open circles and squares show the channels including $\HeLL{6}$
and $\nucLL{\;10}{Be}$, respectively.}
\label{Fig:C12Xi-evn}
\end{figure}

In SDM, the formation probability strongly depends on the $Q$-value
and the number of fragments in the final state
as shown in Fig.~\ref{Fig:C12Xi-evn}.
Since the statistical binary decay favors decays into excited state
having large level densities,
two-body decays with large $Q$-values do not necessarily exhaust
the probabilities.
The formation and decay of $\HeLL{6}$
in the Nagara event~\cite{Nagara} have been uniquely identified to take place as
\begin{align}
\Xi^-+\nuc{12}{C} \to \HeLL{6}+\alpha+t , \ 
\HeLL{6} \to \nucL{5}{He}+p+\pi^-.
\label{Eq:BrNagara}
\end{align}
In the formation of $\nucLL{\;10}{Be}$~\cite{Danysz1963},
the following sequence was found not to be inconsistent
with the Nagara event~\cite{Davis2005,LL_review},
\begin{align}
\Xi^-+\nuc{12}{C} \to \nucLL{\;10}{Be}+d+n ,\ 
\nucLL{\;10}{Be} \to \nucL{9}{Be}^*+p+\pi^-.
\label{Eq:BrDanysz}
\end{align}
These fragmentation branches have relatively large $Q$-values,
$8.4$ and $9.8~\MeV$ in the $\dBLL$ model C,
for channels~\eqref{Eq:BrNagara} and \eqref{Eq:BrDanysz}, respectively.
As a result, the event probability of the decay channel
in Eq.~\eqref{Eq:BrNagara}
(the decay channel in Eq.~\eqref{Eq:BrDanysz})
in SDM is calculated to be $\Pbr=2.6, 1.0$ and $0.9~\%$
($\Pbr=8.8, 3.5$ and $5.0~\%$)
in models A, B and C, respectively.

\begin{table}
\caption{Event probabilities of hypernuclear formation events
including $\HeLL{6}$ and $\nucLL{\;10}{Be}$
from the \DLC\ ($\nucLL{\;13}{B}^*$)
assumed to be formed in the $\Xi^-$ absorption reaction in $\nuc{12}{C}$.
The probabilities ($\Pbr$) are given in \%.}
\label{Tab:C12evn}
\begin{center}
\begin{tabular}{l|rrr}
\hline
\hline
Channel		& \multicolumn{3}{c}{Prob. (\%)}\\
\hline
$\dBLL$ model	& A	& B	& C\\
\hline
$\HeLL{6}+\alpha+t$	&2.6	&1.0	&0.9	\\
$\HeLL{6}+\nuc{7}{Li}$	&0.7	&0.6	&0.5	\\
$\HeLL{6}+\nuc{6}{Li}+n$	&0.4	&0.02	&0.005	\\
$\HeLL{6}+\alpha+d+n$	&0.7	&-	&-	\\
\hline
$\nucLL{\;10}{Be}+d+n$		&8.8	&3.5	&5.0	\\
$\nucLL{\;10}{Be}+t$		&0.7	&0.9	&1.1	\\
$\nucLL{\;10}{Be}+p+n+n$		&9.2	&1.0	&1.9	\\
\hline
\hline
\end{tabular}
\end{center}
\end{table}

We shall now evaluate the expected number of events in the experiment.
In the KEK-E373 experiment, about $10^3$ events of $\Xi^-$ absorption
in emulsion nuclei are analyzed~\cite{E373}.
About half of the $\Xi^-$ absorption events are those with light nuclei,
$\nuc{12}{C}$, $\nuc{14}{N}$ and $\nuc{16}{O}$;
therefore it would be reasonable to assume that about $160$ events
of $\Xi^-$ absorption in $\nuc{12}{C}$ are analyzed.
The formation probability of \DLC, $\nucLL{\;13}{B}^*$,
is about $\PDLC \simeq 30~\%$ in AMD~\cite{Hirata1999}.
Thus the number of decay in a given channel may be given as
$\Nev = 160 \times \PDLC \times \Pbr$ with $\Pbr$ being 
the branching ratio (event probability) in the statistical decay.
For the fragmentation channel in Eq.~\eqref{Eq:BrNagara} (for the formation of $\HeLL{6}$),
the estimated number of events in the E373 experiment is
$\Nev=1.2$, $0.5$ and $0.4$
($\Nev=2.1, 0.8$ and $0.6$)
in models A, B and C, respectively.
Since these numbers are close to unity,
it is not unreasonable that the KEK-E373 experiment
observed one event in the channel of Eq.~\eqref{Eq:BrNagara}.
Since the $\dBLL$ models B and C are more realistic,
we can judge that the KEK-E373 experiment was reasonably lucky
provided that the detection efficiency of $\HeLL{6}$ is high enough.
As for the fragmentation channel in Eq.~\eqref{Eq:BrDanysz} (for the formation of $\nucLL{\;10}{Be}$),
the estimated number of events in the E373 experiment is
$\Nev=4.2$, $1.7$ and $2.4$
($\Nev=14.1, 6.0$ and $8.2$)
in models A, B and C, respectively.
The observation of $\nucLL{\;10}{Be}$ in E373 experiment was reasonable,
as long as the detection efficiency of the sequential weak decay is not small.


Now let us discuss the transport model dependence.
In Fig.~\ref{Fig:C12Xi}, we also show the \DLH\ formation probabilities
obtained by using AMD with additional fluctuations (AMD-QL)
with the statistical decay effects~\cite{Hirata1999}
normalized by $\PDLC=30~\%$.
We show the results with strong fluctuations
at $g_0=0.5$ with $g_0$ being the fluctuation strength parameter,
and a strongly attractive $\Lambda\Lambda$ interaction similar
to the model A is adopted.
Additional fluctuations promote nucleon and $\Lambda$ emissions
in the dynamical stage, then the total \DLC\ formation probability is
found to be around half, $\PDLC(\mathrm{AMD{\hyph}QL})=16~\%$,
compared with the AMD results,
and the total \DLH\ formation probability is also reduced.
By comparison, lighter \DLC\ with smaller excitation energies
are found to be formed in AMD-QL, and the formation probabilities
of some \DLH\ are enhanced.
For example, $\HeLL{6}$ is formed at $\PDLH(\HeLL{6})=2.0~\%$,
which corresponds to $\Pbr(\HeLL{6})\simeq 6.7~\%$ and is larger
than the SDM calculation results.
When normalized by $\PDLC=30~\%$,
the differences of results in AMD-QL with statistical decays
from the SDM results using model A are within a factor of two.
If we assume similar differences in SDM with models B and C,
which are more realistic, expected numbers of events
of $\nucLL{\;10}{Be}$ and $\HeLL{6}$ in KEK-E373 become closer to unity
with additional quantum fluctuation effects in the dynamical stage.
Thus these differences are not negligible but the conclusion
in the previous paragraph, 
the observations of $\nucLL{\;10}{Be}$ and $\HeLL{6}$ in KEK-E373
are reasonable, does not change.

\subsection{$\Xi^-$ absorption in $^{14}\mathrm{N}$}

\begin{figure}[htbp]
\begin{center}
\includegraphics[width=0.7\textwidth]{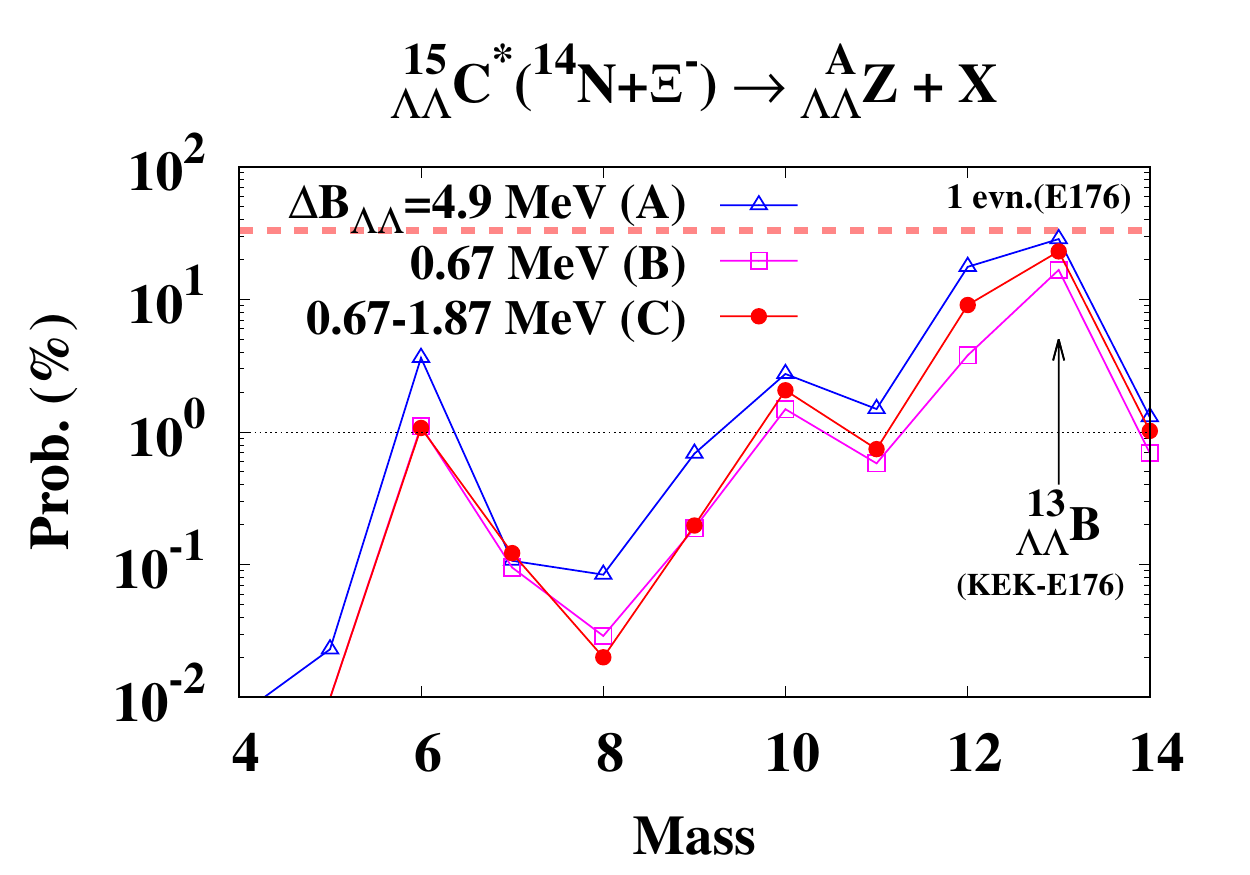}%
\end{center}
\caption{
Double $\Lambda$ hyperfragment mass distribution ($\Pbr$)
in SDM from $\nucLL{\;15}{C}^*$ ($\Xi^-+\nuc{14}{N}$).}
\label{Fig:N14Xi}
\end{figure}

Next we proceed to discuss the $\Xi^-$ absorption in $\nuc{14}{N}$,
where $\nucLL{\;13}{B}$ was observed~\cite{E176}.
In Fig.~\ref{Fig:N14Xi}, we show the \DLH\ mass distribution
in SDM calculation of $\nucLL{\;15}{C}^*$ from $\Xi^-+\nuc{14}{N}$.
As in the case of $\Xi^-+\nuc{12}{C}$,
two nucleon evaporation processes are favored,
and \DLH\ with $A=13$ is most frequently formed;
$\Pbr=(11.2-18.5)~\%$ for $\nucLL{\;13}{B}$
and $\Pbr=(5.4-10.0)~\%$ for $\nucLL{\;13}{C}$.
We also find
$\nucLL{\;12}{B}$ ($\Pbr=(3.4-16.4)~\%$),
$\nucLL{\;10}{Be}$ ($\Pbr=(1.4-2.6)~\%$),
and
$\HeLL{6}$ ($\Pbr=(1.1-3.6)~\%$)
are also frequently formed.
Since the proton separation energy is smaller
and the initial $\nucLL{\;15}{C}^*$ energy is larger for $\nuc{14}{N}$,
the $\dBLL$ model dependence is smaller than other target nuclei.

\begin{figure}[htbp]
\begin{center}
\includegraphics[width=0.7\textwidth]{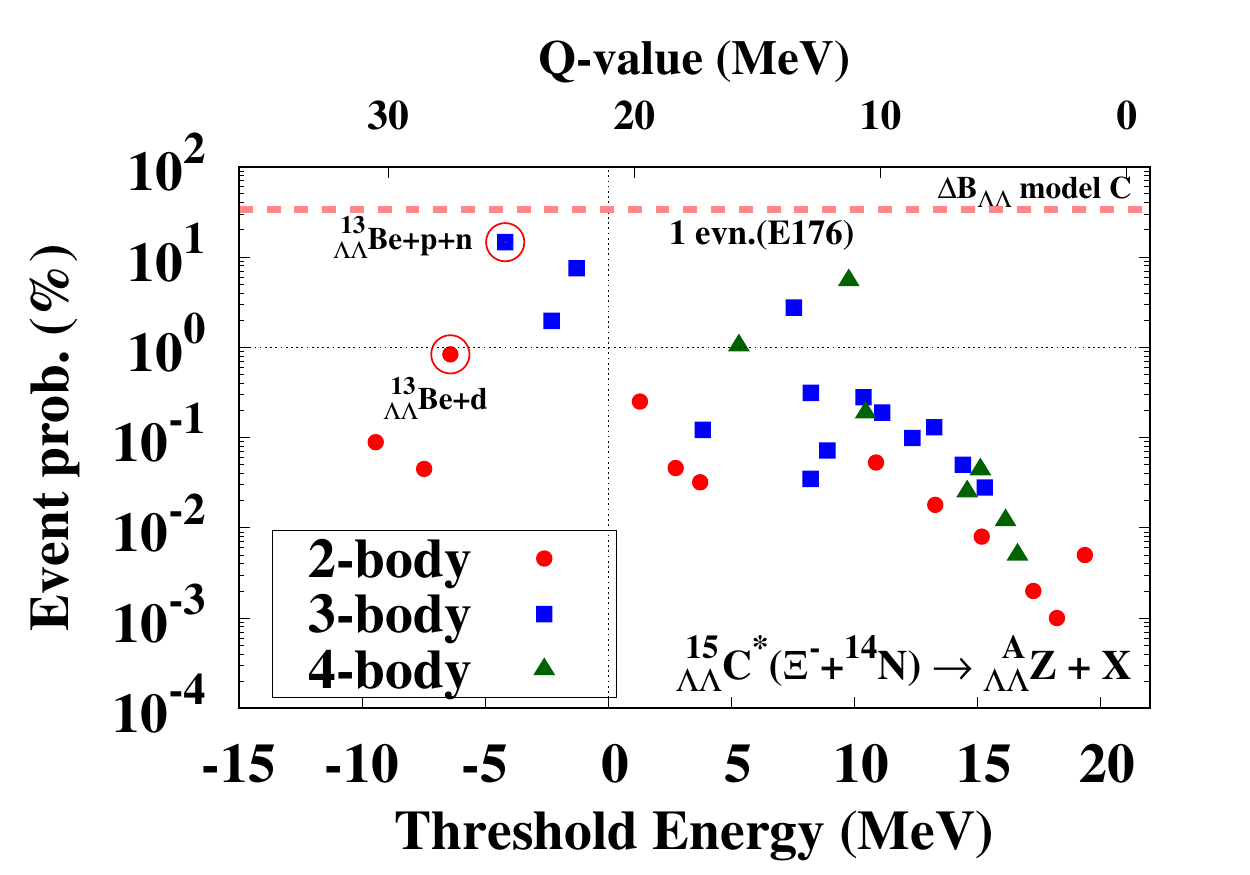}%
\end{center}
\caption{Threshold energy dependence of the event probabilities ($\Pbr$)
in SDM from $\nucLL{\;15}{C}^*$ ($\Xi^-+\nuc{14}{N}$).
Open circles show the channels including $\nucLL{\;13}{B}$.}
\label{Fig:N14Xi-evn}
\end{figure}

In the \DLH\ formation event in E176~\cite{E176},
the following sequence is found to be consistent with the Nagara event~\cite{LL_review},
\begin{align}
\Xi^-+\nuc{14}{N} \to \nucLL{\;13}{B}+p+n ,\ 
\nucLL{\;13}{B} \to \nucL{13}{C}^*+\pi^- .
\label{Eq:BrE176}
\end{align}
As shown in Fig.~\ref{Fig:N14Xi-evn},
this decay channel is found to have the largest event probability;
$\Pbr=17.9$, $10.4$ and $14.6~\%$
($\Pbr=18.5$, $11.2$ and $15.5~\%$)
for the event probability of this channel (formation probability of $\nucLL{\;13}{B}$)
in models A, B and C, respectively.
In the KEK-E176 experiment,
the number of events was estimated to be 
$77.6\pm5.1^{+0.0}_{-12.2}$
for $\Xi^-$ absorption at rest in emulsion~\cite{E176b},
and $31.1 \pm 4.8$
for absorption events on light nuclei~\cite{E176a}.
Then we may roughly estimate the number of $\Xi^-$ absorption events
on $\nuc{14}{N}$ around $10$.
With the \DLC\ formation probability of around $\PDLC \simeq 30~\%$,
the expected number of events in the fragmentation channel in \eqref{Eq:BrE176}
(formation of $\nucLL{\;13}{B}$)
is $\Nev=0.5$, $0.3$ and $0.4$
($\Nev=0.6$, $0.3$ and $0.5$)
in models A, B and C, respectively.
We can judge that the KEK-E176 experiment was also reasonably lucky
provided that the detection efficiency of sequential weak decay of $\nucLL{\;13}{B}$
is high.

\subsection{$\Xi^-$ absorption in $^{16}\mathrm{O}$}

\begin{figure}[htbp]
\begin{center}
\includegraphics[width=0.7\textwidth]{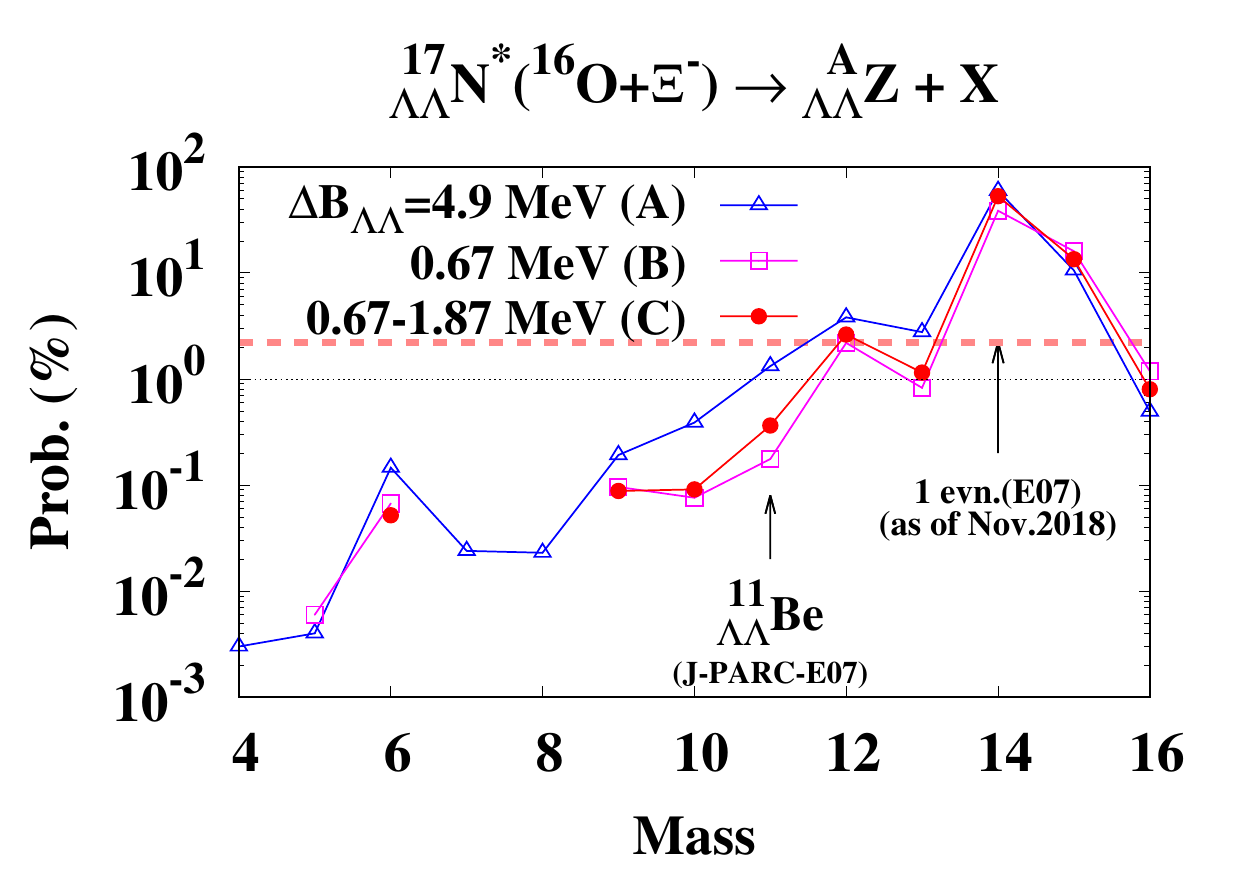}%
\end{center}
\caption{
Double $\Lambda$ hyperfragment mass distribution ($\Pbr$)
in SDM from $\nucLL{\;17}{N}^*$ ($\Xi^-+\nuc{16}{O}$).}
\label{Fig:O16Xi}
\end{figure}

We now discuss the $\Xi^-$ absorption in $\nuc{16}{O}$,
where $\nucLL{}{Be}$ was reported to be observed~\cite{E07}.
In Fig.~\ref{Fig:O16Xi}, we show the \DLH\ mass distribution
in SDM calculation of $\nucLL{\;17}{N}^*$ from $\Xi^-+\nuc{16}{O}$.
The most frequently formed \DLH\ is $\nucLL{\;14}{C}$,
which is a bound state of $\nuc{12}{C}+\Lambda+\Lambda$,
and the formation probability in SDM is
$\Pbr=59.3$, $38.5$ and $52.8~\%$ in models A, B and C, respectively.
The next frequently formed one is $\nucLL{\;15}{C}$,
which is formed via the two nucleon evaporation from $\nucLL{\;17}{N}^*$.
We also find that formation of \DLH\ with $A=7$ and $8$ is strongly suppressed
in the $\dBLL$ models B and C.

\begin{figure}[htbp]
\begin{center}
\includegraphics[width=0.7\textwidth]{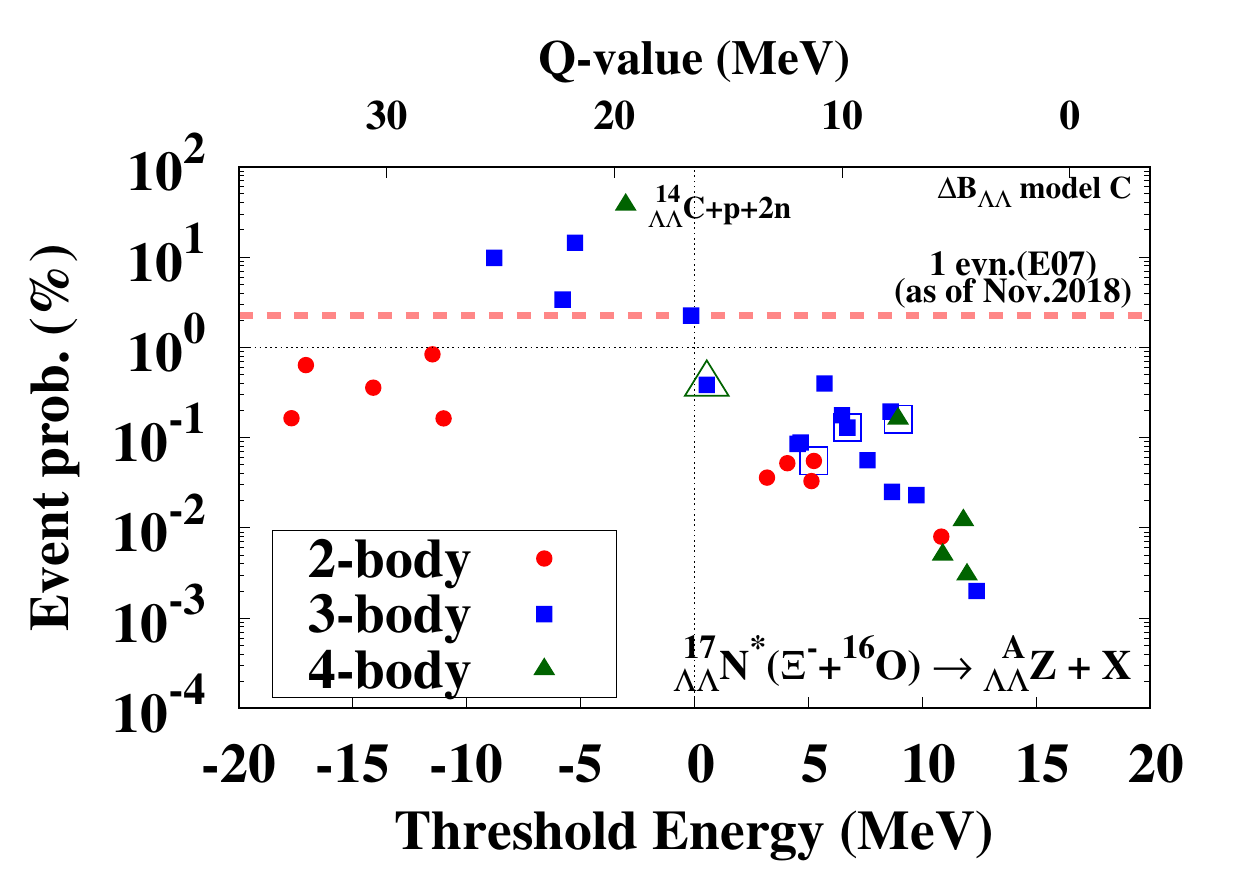}%
\end{center}
\caption{
Threshold energy dependence of the event probabilities ($\Pbr$)
in SDM from $\nucLL{\;17}{N}^*$ ($\Xi^-+\nuc{16}{O}$).
Open circles, squares and triangles show the channels including
$\nucLL{\;10}{Be}$, $\nucLL{\;11}{Be}$ and $\nucLL{\;12}{Be}$,
respectively.}
\label{Fig:O16Xi-evn}
\end{figure}

\begin{table}
\caption{Event probabilities of hypernuclear formation events
including $\nucLL{}{Be}$
from $\nucLL{\;17}{N}^*$ ($\Xi^-+\nuc{16}{O}$).
The probabilities ($\Pbr$) are given in \%.}
\label{Tab:O16evn}
\begin{center}
\begin{tabular}{l|rrr}
\hline
\hline
Channel		& \multicolumn{3}{c}{Prob. (\%)}\\
\hline
$\dBLL$ model	& A	& B	& C\\
\hline
$\nucLL{\;10}{Be}+\nuc{7}{Li}$	&0.07	&0.04	&0.03\\
$\nucLL{\;10}{Be}+\alpha+t$	&0.3	&0.04	&0.06\\
$\nucLL{\;11}{Be}+\nuc{6}{Li}$	&0.08	&0.07	&0.06\\
$\nucLL{\;11}{Be}+\alpha+d$	&0.3	&0.09	&0.1\\
$\nucLL{\;11}{Be}+\alpha+p+n$	&0.7	&0.01	&0.2\\
$\nucLL{\;12}{Be}+\alpha+p$	&0.6	&0.3	&0.4\\
\hline
\hline
\end{tabular}
\end{center}
\end{table}

In the J-PARC E07 experiment,
it is expected to detect 100 \DLH\ formation events
among $10^4$ stopped $\Xi^-$ events.
As of November, 2018, 920 $\Xi^-$ absorption events are analyzed~\cite{E07-QNP2018},
and 8 and 6(+2) double and twin hypernuclear formation events are detected.
Among them, there is one event (Mino event)
where the formed \DLH\ is identified to be $\nucLL{}{Be}$~\cite{E07}.
The event was interpreted as one of the following three candidates,
\begin{align}
&\Xi^-+\nuc{16}{O} \to \nucLL{\;10}{Be}+\alpha+t	 \quad (\dBLL=1.63\pm 0.14~\MeV),\\
&\Xi^-+\nuc{16}{O} \to \nucLL{\;11}{Be}+\alpha+d	 \quad (\dBLL=1.87\pm 0.37~\MeV),\\
&\Xi^-+\nuc{16}{O} \to \nucLL{\;12}{Be}^*+\alpha+p \quad (\dBLL=-2.7\pm 1.0~\MeV). 
\end{align}
Experimentally, the second candidate ($\nucLL{\;11}{Be}$ formation) is considered to be most probable,
and the $\nucLL{\;11}{Be}$ formation probability is reasonably large
($\Pbr=1.1~\%$) in SDM with model A,
as shown in Table~\ref{Tab:Adist}.
By comparison, $\nucLL{\;11}{Be}$ formation is less probable,
$\Pbr=0.2$ and $0.3~\%$, in the models B and C, respectively.
By assuming the number of the $\Xi^-$ absorption events
in $\nuc{16}{O}$ to be around 150
and \DLC\ formation probability of $\PDLC \simeq 30~\%$,
the expected number in the second candidate is $\Nev=0.14, 0.04$ and $0.06$
for models A, B and C, respectively.
In SDM, the third candidate ($\nucLL{\;12}{Be}^*$ formation) is more probable,
the expected number is $\Nev=0.25, 0.15$ and $0.17$ for models A, B and C, respectively.

\section{Double hypernuclear formation from $\Xi^-$ nucleus}
\label{Sec:He6Xi}

\begin{figure}[htbp]
\begin{center}
\includegraphics[width=0.5\textwidth]{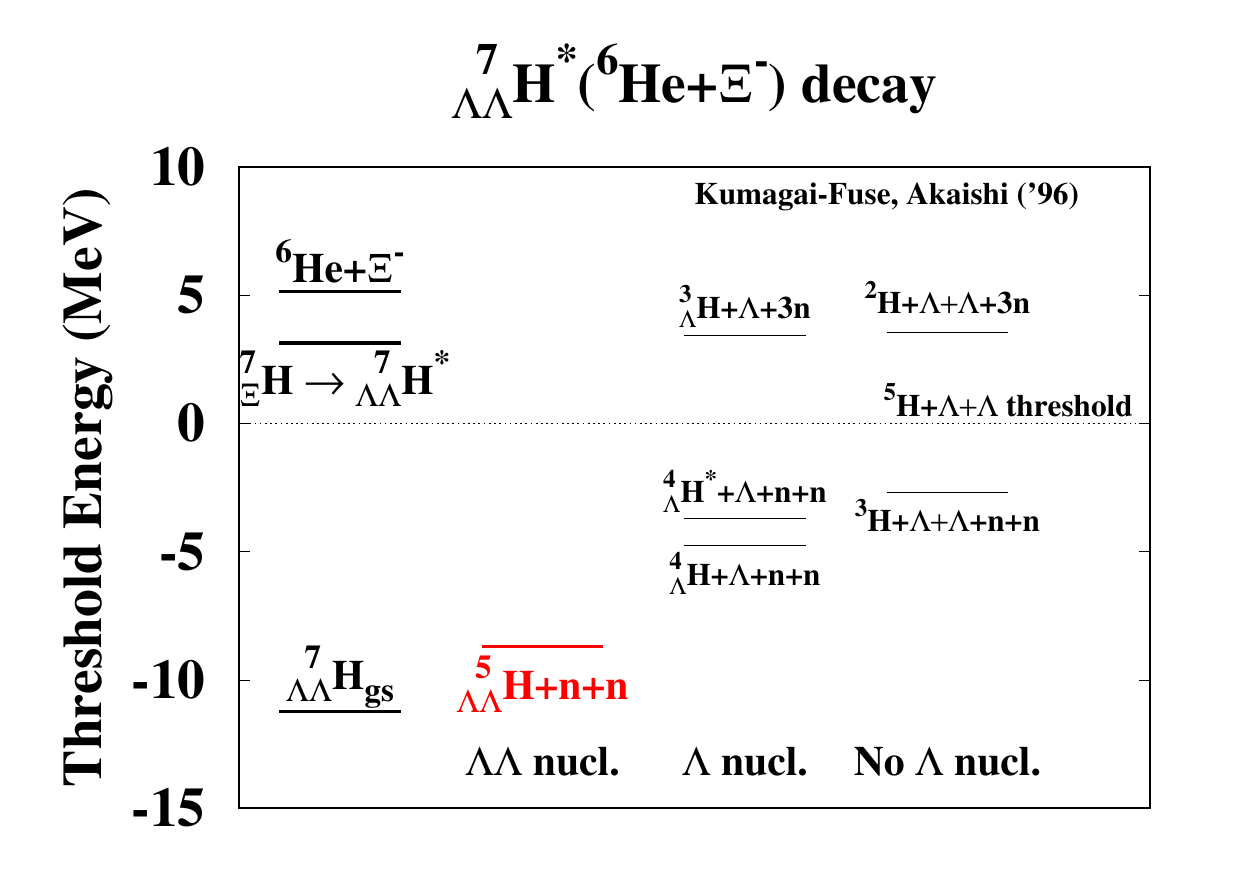}%
\includegraphics[width=0.5\textwidth]{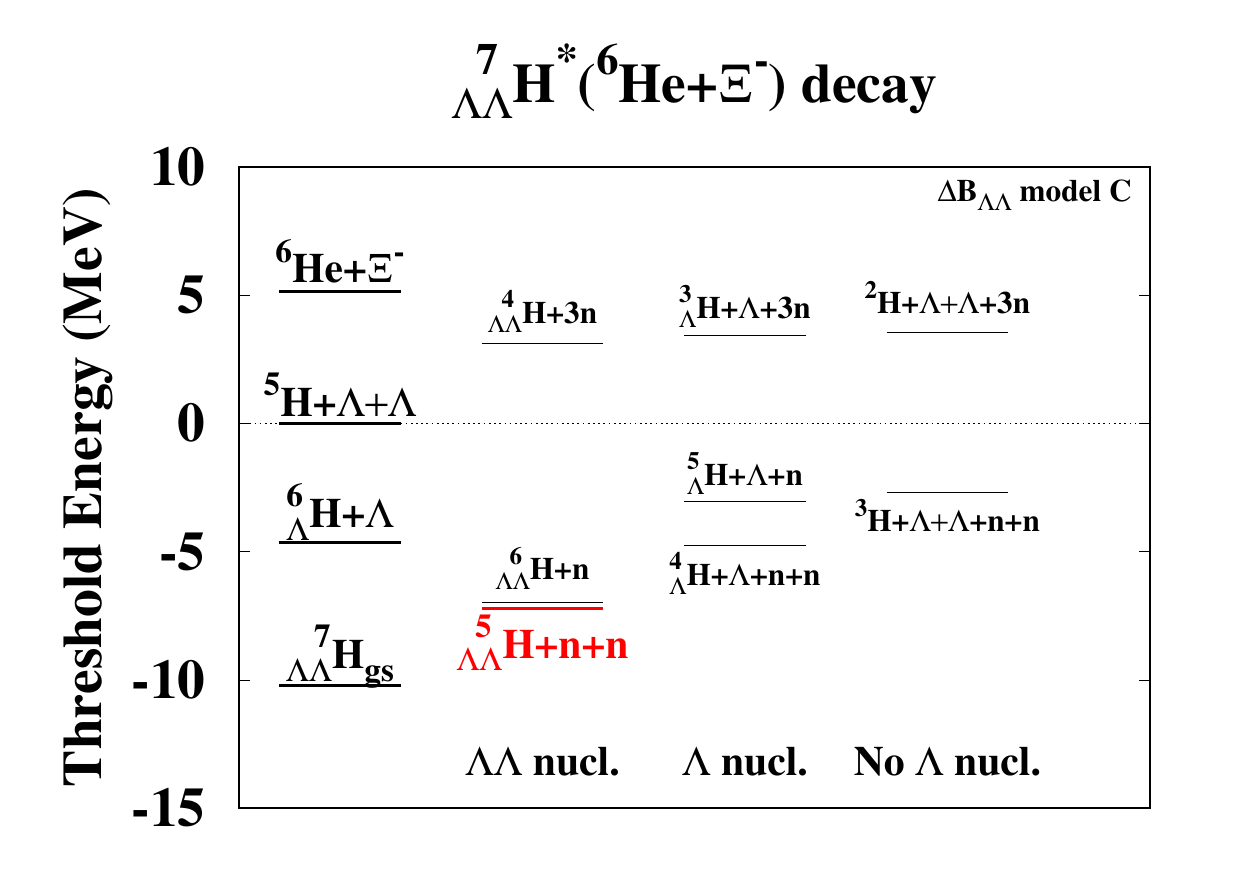}%
\end{center}
\caption{Open channels including \DLH\ from ${}^7_\Xi\mathrm{H}^*$,
in Ref.~\cite{Fuse1996} (left) and in the model C (right).
}
\label{Fig:He6Xi-thr}
\end{figure}

In addition to $\Xi^-$ absorption in nuclei,
conversion of $\Xi$ {\em nuclei} would also make \DLH. 
The $(K^-,K^+)$ reactions on nuclear target populate certain $\Xi$ hypernuclear
states, which become doorway states to \DLH. 
Especially, the $\nuc{7}{Li}(K^-,K^+)$ reaction may produce
the neutron rich $\Xi$ hypernucleus $\nucX{7}{H}=\Xi^-+\nuc{6}{He}$, if exists.
Kumagai-Fuse and Akaishi proposed that the branching ratio
of $\nucX{7}{H}$ conversion to form $\HLL{5}$ is surprisingly large 
to be around $\Pbr=90~\%$~\cite{Fuse1996}.
The first reason of this large branching ratio is the limited decay channels,
\begin{align}
\nucX{7}{H} \to &
\HLL{5}+n+n ,
\nucL{4}{H}+\Lambda+n+n ,
\nucL{4}{H}^*+\Lambda+n+n ,
\nuc{3}{H}+\Lambda+\Lambda+n+n .
\label{Eq:7HXdec}
\end{align}
if the separation energy of $\Xi$ in $\nucX{7}{H}$ is around $2 ~\MeV$
or more.
In the left panel of Fig.~\ref{Fig:He6Xi-thr},
we show the decay channels from $\nucX{7}{H}$.
There are two channels,
$\nucL{3}{H}+\Lambda+3n$ and $\nuc{2}{H}+2\Lambda+3n$,
$1.7$ and $1.6 \MeV$
below the $\Xi^-+\nuc{6}{He}$ threshold,
respectively.
If the binding energy of $\Xi$ in $\nucX{7}{H}$
is larger than $1.7~\MeV$,
$\nucX{7}{H}$ does not decay to these channels.
The second reason of the large branching ratio is in the small $Q$-values.
Since a large part of the released energy in $p\Xi^-\to\LL$ 
($\sim 28.6~\MeV$)
is exhausted in breaking the $\alpha$ cluster in $\nucX{7}{H}$,
the threshold energy of $\Xi^-+\nuc{6}{He}$ is only $5.1~\MeV$
above the $\LL$ emitting threshold ($\nuc{5}{H}+\Lambda+\Lambda$ threshold).
This energy difference is much smaller than in the case of
$\Xi^-$ absorption in 
$\nuc{12}{C}$ (12.7 MeV), $\nuc{14}{N}$ (21.1 MeV) and $\nuc{16}{O}$ (16.5 MeV).
Furthermore, the binding energy of $\Xi^-$ suppresses the $Q$-values.
The $Q$-values in the channels shown in Eq.~\eqref{Eq:7HXdec}
are estimated as $\sim 11, 7, 6$ and $5~\MeV$,
respectively~\cite{Fuse1996}.
These $Q$-values are small and the corresponding "temperatures" are small.
Hence the three-body decays would be favored than four-body or five-body decays.
As a result, the decay process will be dominated
by the $\HLL{5}+2n$ channel
having the largest $Q$-value and three-bodies in the final state.
In Ref.~\cite{Fuse1996}, approximate but explicit calculations were performed
for the decay process of $\nucX{7}{H}$, and a large branching ratio
decaying to $\HLL{5}$ is obtained.
Based on this estimate of high branching ratio to form $\HLL{5}$,
Fujioka \textit{et al.} proposed a new experiment at J-PARC (P75)~\cite{Fujioka2019}.
Since the branching ratio is crucial to determine the feasibility
of the experiment, it is valuable to evaluate the branching ratio
in different approaches.

We shall now discuss the branching ratio of the $\HLL{7}$ decay
to $\HLL{5}+2n$ in SDM described in Sec.~\ref{Sec:SDM}
and used in Sec.~\ref{Sec:results} as it is.
The $\dBLL$ model C is applied as an example.
There are several differences from Ref.~\cite{Fuse1996}.
First, the matrix elements are evaluated statistically as given
in Eq.~\eqref{Eq:SDM}, and excited states in the bound as well as unbound
energy regions are also taken into account.
Second, several other channels are included.
As discussed in Sec.~\ref{Sec:SDM},
nuclei with ground states being resonances are included.
These nuclei include 
$\nuc{5}{H}$, $\nucL{5}{H}$, $\nucL{6}{H}$ and $\HLL{6}$.
By including decay channels containing these nuclei,
$\HLL{7}^*$ can decay in sequential binary decay chains,
for example
$\HLL{7}^* \to \HLL{6}+n \to \HLL{5}+n+n$
and 
$\HLL{7}^* \to \nucL{6}{H}^*+\Lambda
\to \nucL{5}{H}+\Lambda+n \to \nucL{4}{H}+\Lambda+n+n$.
The third point is the binding energy differences.
We adopt the measured and fitted $S_\Lambda$
for single hypernuclei,
and the model C is used for $\dBLL$.
The threshold energies of decay channels in the present treatment are shown
in the right panel of Fig.~\ref{Fig:He6Xi-thr}.

\begin{figure}[htbp]
\begin{center}
\includegraphics[width=0.7\textwidth]{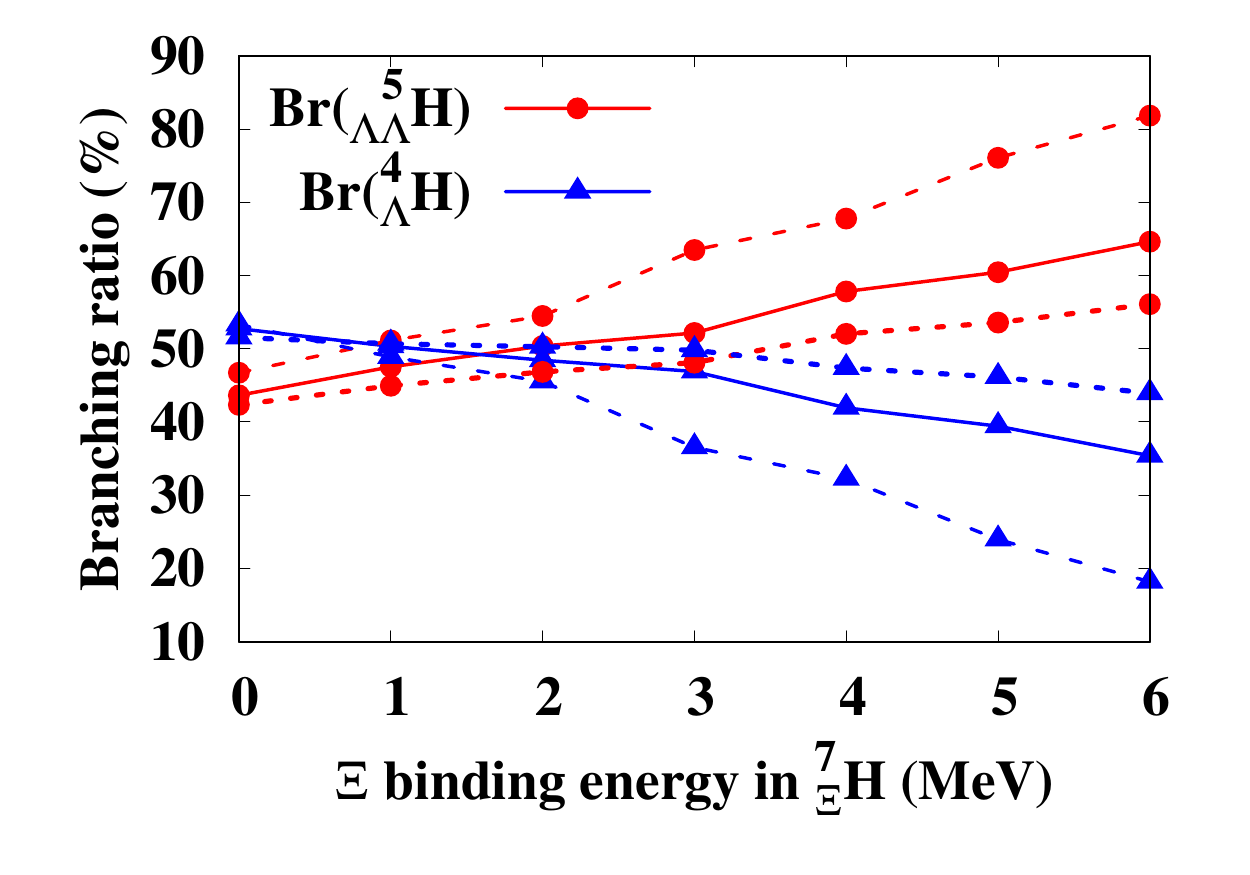}%
\end{center}
\caption{
Branching ratios ($\Pbr$) of
$\HLL{7}\to \HLL{5}+n+n$ (circles)
and
$\HLL{7}\to \nucL{4}{H}+\Lambda+n+n$ (triangles)
in SDM as functions of the binding energy of $\Xi^-$ in $\nucX{7}{H}$.
Solid (dotted) lines show the results in the $\dBLL$ model C
with (without) dineutron emission.
Dashed lines show the results without considering particle unstable nuclei
in the decay processes and with the channels and binding energies
used in \cite{Fuse1996}.}
\label{Fig:He6Xi-brat}
\end{figure}

In Fig.~\ref{Fig:He6Xi-brat}, we show
the branching ratios of
$\HLL{7}\to \HLL{5}+n+n$ ($\Pbr(\HLL{5})$)
and
$\HLL{7}\to \nucL{4}{H}+\Lambda+n+n$ ($\Pbr(\nucL{4}{H})$)
obtained in the SDM calculation
using the $\dBLL$ model C
as functions of the binding energy of $\Xi^-$ in $\nucX{7}{H}$
(solid curves).
The $\Xi^-$ binding energy in $\nucX{15}{C}$ was evaluated
from the twin hypernuclear formation in the Kiso event,
$\Xi^-+\nuc{14}{N} \to \nucX{15}{C} \to \nucL{10}{Be}^{(*)}+\nucL{5}{He}$,
as $B_\Xi=3.87 \pm 0.21~\MeV$
or $1.03 \pm 0.18~\MeV$~\cite{Nakazawa:2015joa,LL_review}.
These two values are for decays to the ground state and the excited state
of $\nucL{10}{Be}$.
The binding energy would be smaller in $\nucX{7}{H}$,
so the binding energy region of $0\leq B_\Xi \leq 4~\MeV$
would be enough.

The branching ratio $\Pbr(\HLL{5})$ increases
with increasing $B_\Xi$, as expected,
and takes values between $\Pbr=43.6~\%$ ($B_\Xi=0$)
and $\Pbr=57.8~\%$ ($B_\Xi=4~\MeV$)
in the $B_\Xi$ region of interest.
These values are smaller than those in Ref.~\cite{Fuse1996},
but still a large branching ratio around $\Pbr=50~\%$ seems
to be probable in SDM.

Let us discuss the difference of the branching ratio
in the present study and in Ref.~\cite{Fuse1996}.
We show the SDM results without dineutron emission
by dotted lines in Fig.~\ref{Fig:He6Xi-brat}.
The model C is used for $\dBLL$.
In this case,
$\nucL{6}{H} (B(\nucL{6}{H})=10.44~\MeV)$
is assumed to decay into 
$\nucL{4}{H} (B(\nucL{4}{H})=10.52~\MeV)$
in the final state, while the one neutron emission into
$\nucL{5}{H} (B(\nucL{5}{H})=8.83~\MeV)$
is not allowed energetically.
The branching ratio to form $\HLL{5}$ takes
$\Pbr(\HLL{5})=42.3~\%$ ($B_\Xi=0$)
and $52.0~\%$ ($B_\Xi=4~\MeV$).
These values are slightly smaller than those with dineutron emission.
This is because
the direct three-body decay of $\HLL{7}^*$ into $\HLL{5}+2n$
simulated by dineutron emission is suppressed,
while the effect of dineutron emission is not significant
and hypernuclear formation is dominated by binary decay sequences 
such as
$\HLL{7}^* \to \HLL{6}^{(*)}+n \to \HLL{5}+2n$
and 
$\HLL{7}^* \to \nucL{6}{H}^{(*)}+\Lambda \to \nucL{5}{H}+\Lambda+n$.
Since the $Q$-values are small, it is reasonable to expect that
two-body (binary) decays are expected to be favored
than three-body or four-body decays.
We also show the branching ratios in SDM 
with the $\dBLL$ values and channels used in Ref.~\cite{Fuse1996}
by dashed lines in Fig.~\ref{Fig:He6Xi-brat}.
Here we adopt $\dBLL=1.92~\MeV$
and ignore unstable hypernuclei with respect to particle decay,
$\nucL{5}{H}$, $\nucL{6}{H}$ and $\HLL{6}$, in the decay processes.
The branching ratio to $\HLL{5}$ takes larger values than 
those with particle unstable nuclei, and takes values of
$\Pbr(\HLL{5})=46.7~\%$ ($B_\Xi=0$)
and 
$67.8~\%$ ($B_\Xi=4~\MeV$).
%
These values are still smaller than those in Ref.~\cite{Fuse1996}.
Therefore the larger branching ratio around $\Pbr=90~\%$ in Ref.~\cite{Fuse1996}
is found to be a result of an explicit evaluation of the transition width
in a specific model treatment in addition to 
the kinematical reasons such as the limited number of decay channels
and ignoring particle unstable states during the decay processes.
For a more serious estimate, we need calculations
with updated $\Xi^-N$ and $\LL$ interactions
and experimental confirmation of the three-body decay width
in the same theoretical treatment.

It should be noted that the formation of \DLC\ and its statistical decay
may be a less reliable picture of hypernuclear formation
for lighter nuclear targets.
For example, while AMD+SDM roughly explains
the $\nucL{4}{H}$ formation probability
from the $K^-$ absorption reaction at rest in $\nuc{12}{C}$ and $\nuc{16}{O}$,
the combined framework underestimates~\cite{Nara1995} the probability for
$\nuc{7}{Li}$ and $\nuc{9}{Be}$ targets~\cite{Tamura:1988kk}.
These light target nuclei, $\nuc{7}{Li}$ and $\nuc{9}{Be}$,
have cluster structure in the ground states and easily dissociate to fragments
after the conversion process of $K^-N\to\pi\Lambda$.
Then \DLH\ formation may take place without going through \DLC.

\section{Summary and discussion}
\label{Sec:Summary}

We have investigated the double $\Lambda$ hypernuclear (\DLH) formation
from the double $\Lambda$ compound nuclei (\DLC),
$\nucLL{\;13}{B}^*$, $\nucLL{\;15}{C}^*$, $\nucLL{\;17}{N}^*$ and $\HLL{7}^*$
in the statistical decay model (SDM).
The first three compound nuclei would be formed in the $\Xi^-$ absorption
at rest in light nuclear target in the emulsion,
$\Xi^-+\nuc{12}{C}\to\nucLL{\;13}{B}^*$,
$\Xi^-+\nuc{14}{N}\to\nucLL{\;15}{C}^*$,
and 
$\Xi^-+\nuc{16}{O}\to\nucLL{\;17}{N}^*$,
and the last one would be formed from the $\Xi$-hypernuclei,
$\nuc{7}{Li}(K^-,K^+)\nucX{7}{H}_\mathrm{gs}$
followed by $\nucX{7}{H}_\mathrm{gs} \to \HLL{7}^*$.
SDM has been demonstrated to work well in describing
fragment formation processes,
especially when combined with the transport model calculation
to populate the compound nuclei and the excitation energies.
In the antisymmetrized molecular dynamics (AMD) calculations
of $\Xi^-$ absorption in $\nuc{12}{C}$,
the formed 
\DLC\ 
are dominated by 
$\nucLL{\;13}{B}^*$ and its probability is around $\PDLC \simeq 30~\%$.
Assuming that this mechanism also applies to the $\Xi^-$ absorption
in other target nuclei,
we have evaluated the formation probabilities of \DLH.
We have also applied the same SDM to evaluate
the branching ratio of the $\Xi$ hypernucleus $\nucX{7}{H}$
to form $\HLL{5}$.

We have examined the target and the $\LL$ bond energy ($\dBLL$)
dependence of the \DLH\ formation probabilities,
and the event probabilities are also examined 
for the channels in which \DLH\ have been observed in experiments.
The bond energy is given as
$\dBLL=4.9~\MeV$ (model A),
$\dBLL=0.67~\MeV$ (model B),
and 
$\dBLL(A=6)=0.67~\MeV$ and $\dBLL(A=11)=1.87~\MeV$ (model C).
In models A and B, $\dBLL$ is assumed to be independent of the hypernuclei,
and in the model C, $\dBLL$ is assumed to be
a linear function of the \DLH\ mass number.
Models B and C are consistent with the Nagara event result.

In the $\Xi^-$ absorption in $\nuc{12}{C}$,
$\HeLL{6}$ and $\nucLL{\;10}{Be}$
are observed in the fragmentation channel of
$\Xi^-+\nuc{12}{C} \to \HeLL{6}+\alpha+t$
and 
$\Xi^-+\nuc{12}{C} \to \nucLL{\;10}{Be}+d+n$, respectively.
In SDM, these channels are found to have relatively large probabilities,
and the expected numbers of events are consistent with unity
within a factor of three.
For $\nuc{14}{N}$ target, $\nucLL{\;13}{B}$ is observed in the channel
$\Xi^-+\nuc{14}{N} \to \nucLL{\;13}{B}+p+n$.
The SDM calculation shows that this channel has the largest event probability
among other fragmentation channels,
and the expected number of events is $(0.3-0.4)$
in $\dBLL$ models B and C.
For $^{16}\mathrm{O}$ target, $\nucLL{}{Be}$ is observed.
The mass number and channels are not uniquely identified,
but the experimentally most probable channel is
$\Xi^-+\nuc{16}{O} \to \nucLL{\;11}{Be}+\alpha+d$.
This channel is found to have small event probability of $\Pbr=(0.2-0.3)~\%$
and the expected number of events is $(0.04-0.06)$
in $\dBLL$ models B and C.
If the observation was not accidental,
we may need other mechanisms than the statistical decay of \DLC.

The $\dBLL$ model dependence of the \DLH\ formation probabilities
is not significant in the main decay channels, 
where a few nucleons are evaporated;
two nucleon emission for $\nuc{12}{C}$ and $\nuc{14}{N}$ targets
and three nucleon emission for $\nuc{16}{O}$,
as shown in Figs.~\ref{Fig:C12Xi}, \ref{Fig:N14Xi} and \ref{Fig:O16Xi}.
One nucleon emission to excited levels is favored
in each step of the statistical binary decay 
because of the large level densities in the daughter nuclei,
$\rho \propto A^{-5/3}e^{2\sqrt{aE^*}}$
with $a$ being the level density parameter,
$a\simeq A/8\ \MeV^{-1}$~\cite{SDM,FaiRandrup1982}.
Since the mass number dependence of $\dBLL$ is assumed to be small,
the $\dBLL$ difference results in the shift of the excitation energies
of the parent and daughter nuclei simultaneously
and does not affect the decay width of one nucleon emission strongly.
By comparison, lighter \DLH\ formation depends more strongly on $\dBLL$.
Since the initial energy is fixed in $\Xi^-$ absorption in nuclei,
the $Q$-value difference of around $4\ \MeV$ in emitting light \DLH\ 
is significant in the total released energy of $(10-30)\ \MeV$
in the fragmentation.
Transport model dependence is also discussed
via the comparison with the results
in AMD with quantum fluctuation effects~\cite{Hirata1999}, 
and a factor two difference may appear in the \DLH\ formation probabilities.

We have also discussed the branching ratio to form $\HLL{5}$
from a $\Xi$ hypernucleus $\nucX{7}{H}$,
which can be formed in the $\nuc{7}{Li}(K^-,K^+)$ reaction
and is assumed to be converted to a \DLC, $\HLL{7}^*$.
It was proposed that the branching ratio would be
around $\Pbr=90~\%$~\cite{Fuse1996}.
In SDM calculations, the branching ratio is found to be $\Pbr=(40-60)~\%$
in the $\Xi$ binding energy range of $(0-4)\ \MeV$.
This branching ratio is still large but lower than that in Ref.~\cite{Fuse1996}.
The difference seems to come from the method to evaluate the decay width;
explicit few-body calculations in \cite{Fuse1996}
and the statistical assumptions in the present work.

The theoretical framework adopted in the present work
may be too much simplified to describe realistic \DLH\ formation processes,
and we need improvements in the theoretical frameworks
for more quantitative estimates of the formation probabilities of \DLH\ 
from $\Xi^-$ absorption reactions at rest and $\Xi$ hypernuclei.
First, the initial \DLC\ formation probabilities need to be evaluated
with updated $\dBLL$ values consistent with the Nagara event
and for $\nuc{14}{C}$, $\nuc{14}{N}$ and $\nuc{16}{O}$ target nuclei
in a consistent way.
We have adopted the formation probability of \DLC\ 
in the initial preequilibrium stage of around $\PDLC \simeq 30~\%$,
based on the antisymmetrized molecular dynamics (AMD)
calculation of $\Xi^-$ absorption at rest in $\nuc{12}{C}$~\cite{Hirata1999}. 
This probability is obtained with a strongly attractive $\LL$ interaction,
which gives the $\LL$ bond energy $\dBLL=4.9\ \MeV$ in $\nucLL{\;13}{B}$,
and $\nuc{14}{N}$ and $\nuc{16}{O}$ targets were not considered.
In heavier targets ($\nuc{14}{N}$ and $\nuc{16}{O}$),
the $\Lambda$ trapping probabilities are found to be larger
in the case of $K^-$ absorption at rest~\cite{Nara1995}
and then similar \DLC\ formation probabilities or more are also expected,
while this expectation may be too optimistic and is premature.
Second, we need more care to discuss hypernuclear formation
in reactions on light nuclear targets such as $\nuc{7}{Li}$ and $\nuc{9}{Be}$.
When the ground state of the target has a developed cluster structure,
direct reaction processes are found to be more important
than in reactions on C, N and O targets~\cite{Nara1995}.
This may also applies to the decay of light $\Xi^-$ nucleus
such as $\nucX{7}{H}$ as discussed in Sec.~\ref{Sec:He6Xi}.
Thirdly, we may need other quantum effects or multifragmentation mechanisms
to understand the \DLH\ formation consistently with
the twin hypernuclear formation,
where two single $\Lambda$ hypernuclei are formed.
The twin hypernuclear formation probability is known to be comparable
to that of \DLH, but SDM predicts much lower probability
of twin hypernuclear formation.
In Ref.~\cite{Hirata1999}, quantum fluctuation effects are considered
and are found to promote twin hypernuclear formation,
but the twin hypernuclear formation probability is still underestimated.
One of the possible mechanisms for producing twin single hypernuclei efficiently
is to go through the resonance states of two single hypernuclei
around the threshold energy of $\Xi^-$ and target nuclei~\cite{Yamada1997}.
This mechanism seems plausible, since
resonance states tend to appear around the threshold,
and there are several twin hypernuclear channels
around the $\Xi^-$-target threshold.
Another candidate mechanism is the multifragmentation,
simultaneous decay to three or more fragments.
The excitation energy of $E^*=(30-50)\ \MeV$ corresponds to 
the temperature of $T \simeq \sqrt{E^*/a} \sim 5\ \MeV$
and multifragmentation may be relevant at this temperature.
Actually,
canonical and microcanonical statistical fragmentation models are applied
to hypernuclear formation in Refs.~\cite{Yamamoto1994,Lorente2011}.
Thus it is desired to examine \DLH\ formation probabilities
in multifragmentation models.

\section*{Acknowledgments}
The authors would like to thank
Hiroyuki Fujioka, Tomokazu Fukuda and Emiko Hiyama
for giving us motivation for this work,
Hirokazu Tamura, Toshiyuki Gogami,
for telling us the recalibration results in SKS data,
and J{\o}rgen Randrup
for useful discussions and careful reading of our manuscript.
This work is supported in part by the Grants-in-Aid for Scientific Research
from JSPS (Nos. 
19H05151 
and
19H01898), 
and by the Yukawa International Program for Quark-hadron Sciences (YIPQS).

\bibliographystyle{ptephy}

\begin{thebibliography}{99}
\bibitem{Danysz1963}
  M. Danysz {\it et al.},
  Phys. Rev. Lett.  {\bf 11} (1963), 29;
  Nucl. Phys.  {\bf 49} (1963), 121.

\bibitem{E176}
  S. Aoki {\it et al.},
  Prog. Theor. Phys. {\bf 85} (1991), 1287.

\bibitem{Nagara}
  H. Takahashi {\it et al.},
  Phys. Rev. Lett.  {\bf 87} (2001), 212502.

\bibitem{E373}
  J. K. Ahn {\it et al.} [E373 (KEK-PS) Collaboration],
  Phys. Rev. C {\bf 88} (2013), 014003.

\bibitem{E07}
  H.~Ekawa {\it et al.},
  PTEP {\bf 2019} (2019), 021D02.

\bibitem{Hiyama2010}
  E. Hiyama, M. Kamimura, Y. Yamamoto and T. Motoba,
  Phys. Rev. Lett.  {\bf 104} (2010), 212502.

\bibitem{FG2002}
  I. N. Filikhin and A. Gal,
  Nucl. Phys. A {\bf 707} (2002), 491.

\bibitem{ESC08}
  T.~A.~Rijken, M.~M.~Nagels and Y.~Yamamoto,
  Prog.\ Theor.\ Phys.\ Suppl.\  {\bf 185} (2010) 14.

\bibitem{fss2}
  Y.~Fujiwara, Y.~Suzuki and C.~Nakamoto,
  Prog.\ Part.\ Nucl.\ Phys.\  {\bf 58} (2007) 439.


\bibitem{Sasaki:2015ifa}
  K.~Sasaki {\it et al.} [HAL QCD Collaboration],
  PTEP {\bf 2015} (2015), 113B01.

\bibitem{CorrExp}
  L. Adamczyk {\it et al.} [STAR Collaboration],
  Phys. Rev. Lett.  {\bf 114} (2015), 022301;
  S. Acharya {\it et al.} [ALICE Collaboration],
  Phys. Rev. C {\bf 99} (2019), 024001;
  J. Adam {\it et al.} [STAR Collaboration],
  Phys. Lett. B {\bf 790} (2019), 490;
  S. Acharya {\it et al.} [ALICE Collaboration],
  Phys. Rev. Lett. {\bf 123} (2019), 112002.


\bibitem{CorrTheory}
  K. Morita, T. Furumoto and A. Ohnishi,
  Phys. Rev. C {\bf 91} (2015),  024916;
  K. Morita, A. Ohnishi, F. Etminan and T. Hatsuda,
  Phys. Rev. C {\bf 94} (2016), 031901;
  A. Ohnishi, K. Morita, K. Miyahara and T. Hyodo,
  Nucl. Phys. A {\bf 954} (2016), 294;
  S. Cho {\it et al.} [ExHIC Collaboration],
  Prog. Part. Nucl. Phys.  {\bf 95} (2017), 279;
  T. Hatsuda, K. Morita, A. Ohnishi and K. Sasaki,
  Nucl. Phys. A {\bf 967} (2017), 856;
  D. L. Mihaylov, V. Mantovani Sarti, O. W. Arnold, L. Fabbietti, B. Hohlweger and A. M. Mathis,
  Eur. Phys. J. C {\bf 78} (2018), 394;
  J. Haidenbauer,
  Nucl. Phys. A {\bf 981} (2019), 1;
  K. Morita, S. Gongyo, T. Hatsuda, T. Hyodo, Y. Kamiya and A. Ohnishi,
  Phys. Rev. C {\bf 101} (2020), 015201.

\bibitem{HypEOS}
  C. Ishizuka, A. Ohnishi, K. Tsubakihara, K. Sumiyoshi and S. Yamada,
  J. Phys. G {\bf 35} (2008) 085201;
  K. Tsubakihara, H. Maekawa, H. Matsumiya and A. Ohnishi,
  Phys. Rev. C {\bf 81} (2010) 065206.

\bibitem{MassiveNS}
  P. Demorest, T. Pennucci, S. Ransom, M. Roberts and J. Hessels,
  Nature {\bf 467} (2010), 1081;
  J. Antoniadis {\it et al.},
  Science {\bf 340} (2013), 6131;
  H. T. Cromartie {\it et al.},
  arXiv:1904.06759 [astro-ph.HE].

\bibitem{3BR}
  S. Nishizaki, T. Takatsuka and Y. Yamamoto,
  Prog. Theor. Phys.  {\bf 108} (2002), 703;
  J. Rikovska-Stone, P. A. M. Guichon, H. H. Matevosyan and A. W. Thomas,
  Nucl. Phys. A {\bf 792} (2007), 341;
  T. Miyatsu, S. Yamamuro and K. Nakazato,
  Astrophys. J.  {\bf 777} (2013), 4;
  D. Lonardoni, A. Lovato, S. Gandolfi and F. Pederiva,
  Phys. Rev. Lett.  {\bf 114} (2015),  092301;
  H. Togashi, E. Hiyama, Y. Yamamoto and M. Takano,
  Phys. Rev. C {\bf 93} (2016), 035808.

\bibitem{Tsubakihara2013}
  K. Tsubakihara and A. Ohnishi,
  Nucl. Phys. A {\bf 914} (2013), 438.

\bibitem{PANDA}
  A.~Sanchez Lorente [Panda Collaboration],
  Hyperfine Interact.\  {\bf 229} (2014) no.1-3,  45.

\bibitem{DLC}
  T. Yamazaki and H. Tamura,
  Czech. J. Phys.  {\bf 42} (1992), 1137.

\bibitem{HIC}
  J.~Steinheimer, K.~Gudima, A.~Botvina, I.~Mishustin, M.~Bleicher and H.~Stocker,
  Phys.\ Lett.\ B {\bf 714} (2012) 85.


\bibitem{Prowse1966} 
  D. J. Prowse,
  Phys. Rev. Lett.  {\bf 17} (1966), 782.
  doi:10.1103/PhysRevLett.17.782

\bibitem{Dalitz1989} 
  R. H. Dalitz, D. H. Davis, P. H. Fowler, A. Montwill, J. Pniewski and J. A. Zakrzewski,
  Proc. Roy. Soc. Lond. A {\bf 426} (1989), 1.

\bibitem{Yamamoto1994}
  Y. Yamamoto, M. Sano and M. Wakai,
  Prog. Theor. Phys. Suppl.  {\bf 117} (1994), 265.

\bibitem{Hirata1999}
  Y. Hirata, Y. Nara, A. Ohnishi, T. Harada and J. Randrup,
  Prog. Theor. Phys.  {\bf 102} (1999), 89.

\bibitem{Lorente2011}
  A. S. Lorente, A. S. Botvina and J. Pochodzalla,
  Phys. Lett. B {\bf 697} (2011), 222.

\bibitem{Yamada1997}
  T. Yamada and K. Ikeda,
  Phys. Rev. C {\bf 56} (1997), 3216.
  doi:10.1103/PhysRevC.56.3216

\bibitem{SDM}
  F. P{\"u}hlhofer,
  Nucl. Phys. A {\bf 280} (1977), 267.

\bibitem{FaiRandrup1982}
  G. I. Fai and J. Randrup,
  Nucl. Phys. A {\bf 381} (1982), 557.

\bibitem{MD_SDM}
  A. Ono, H. Horiuchi, T. Maruyama and A. Ohnishi,
  Prog. Theor. Phys.  {\bf 87} (1992), 1185;
  Phys. Rev. Lett.  {\bf 68} (1992), 2898;
  T. Maruyama, A. Ono, A. Ohnishi and H. Horiuchi,
  Prog. Theor. Phys.  {\bf 87} (1992), 1367.

\bibitem{Nara1995}
  Y. Nara, A. Ohnishi and T. Harada,
  Phys. Lett. B {\bf 346} (1995), 217.

\bibitem{BMZ}
  H. Bando, T. Motoba and J. Zofka,
  Int. J. Mod. Phys. A {\bf 5} (1990), 4021.
\bibitem{HT}
  O. Hashimoto and H. Tamura,
  Prog. Part. Nucl. Phys.  {\bf 57} (2006), 564.

\bibitem{Gogami2016}
  T.~Gogami {\it et al.},
  Phys.\ Rev.\ C {\bf 93} (2016) no.3,  034314.
\bibitem{E176a}
K. Nakazawa, in {\em Proc. of the 23rd INS International Symposium
on Nuclear and Particle Physics with Meson Beams in the 1 GeV/c Region},
Tokyo, Japan, Mar. 15-18, 1995, ed. S. Sugimoto and O. Hashimoto
(Universal Academy Press, Inc., Tokyo, Japan, 1995), p. 261.

\bibitem{E176b}
  S. Aoki {\it et al.} [KEK E176 Collaboration],
  Nucl. Phys. A {\bf 828} (2009), 191.

\bibitem{E373trap}
  A.~M.~M.~Theint {\it et al.},
  PTEP {\bf 2019} (2019), 021D01.


\bibitem{Davis2005}
  D. H. Davis,
  Nucl. Phys. A {\bf 754} (2005), 3.

\bibitem{LL_review}
   E. Hiyama and K. Nakazawa,
   Annu. Rev. Nucl. Part. Sci. \textbf{68} (2018), 131.


\bibitem{E07-QNP2018}
J. Yoshida {\it et al.} [J-PARC E07 Collaboration],
JPS Conf. Proc. {\bf 26} (2019), 023006. 



\bibitem{Fuse1996}
  I. Kumagai-Fuse and Y. Akaishi,
  Phys. Rev. C {\bf 54} (1996), R24.

\bibitem{Fujioka2019}
  H. Fujioka, T. Fukuda, E. Hiyama, T. Motoba, T. Nagae, S. Nagao and T. Takahashi,
  AIP Conf. Proc.  {\bf 2130} (2019),  040002.

\bibitem{Nakazawa:2015joa}
  K. Nakazawa {\it et al.},
  PTEP {\bf 2015} (2015), 033D02.

\bibitem{Tamura:1988kk}
  H. Tamura {\it et al.},
  Phys. Rev. C {\bf 40} (1989), 479(R).




\end{thebibliography}

\end{document}